\documentclass{aa}  
\usepackage{bm} %bold math (vectors)
\usepackage{txfonts}
\usepackage{natbib} 
\usepackage{nicefrac}
\usepackage{graphicx}
\usepackage[table,  dvipsnames]{xcolor} %for author comment colors and that table
\usepackage{nicefrac} %for some nice inline fractions
\usepackage{hyperref} % To add links to your PDF file, use the package "hyperref"           
\usepackage{placeins}
\usepackage{multicol}

\hypersetup{colorlinks = {true},
            linkcolor = [rgb]{0,0.35,0.7},
            citecolor = [rgb]{0,0.35,0.7},
            filecolor = [rgb]{0.61,0,0},
            urlcolor = [rgb]{0.61,0,0},
           }

\graphicspath{ {} } %place figures there

\usepackage{soul}

\newcommand{\editor}[1]{#1}
\newcommand{\referee}[1]{#1}

\newcommand{\DP}[2]{\frac{\partial{#1}}{\partial{#2}}}

\newcommand{\Int}{\int\limits}
\newcommand{\rbeg}{r_\text{beg}}

\newcommand{\phistar}{\Phi_\star}
\newcommand{\Mstar}{M_\star}
\newcommand{\Msun}{\text{M}_{\odot}}
\newcommand{\Rsun}{\text{R}_{\odot}}
\newcommand{\Fstar}{F_\star}
\newcommand{\Rstar}{R_\star}
\newcommand{\Tstar}{T_\star}
\newcommand{\Lstar}{L_\star}

\newcommand{\Etot}{E_\text{tot}}
\newcommand{\Erad}{E_\text{rad}}
\newcommand{\Frad}{F_\text{rad}}
\newcommand{\bmFrad}{\bm{F}_\text{rad}}
\newcommand{\arad}{a_\text{R}}

\newcommand{\kp}{\kappa_\text{P}}
\newcommand{\kr}{\kappa_\text{R}}
\newcommand{\knu}{\kappa_\nu}

\newcommand{\G}{\text{G}}
\newcommand{\sigsb}{\sigma_\text{SB}}

\newcommand{\cs}{c_\mathrm{s}}

\newcommand{\OmegaK}{\Omega_\mathrm{K}}

\newcommand{\cv}{c_\mathrm{v}}

\newcommand{\St}{\text{St}}
\newcommand{\edg}{\epsilon_\text{dg}}

\authorrunning{Sudarshan et al.}
\titlerunning{flared-staircase disk structures}

\begin{document} 
\title{Starlight-driven flared-staircase geometry in radiation hydrodynamic models of protoplanetary disks}

\author{Prakruti Sudarshan
        \inst{\ref{inst1}, \ref{inst2}} \and
        Mario Flock\inst{\ref{inst1}}, 
        Alexandros Ziampras\inst{\ref{inst3}, \ref{inst1}}, David Melon Fuksman\inst{\ref{inst1}} \and Tilman Birnstiel\inst{\ref{inst3}, \ref{inst4}}
        }
\institute{
           Max-Planck-Institut für Astronomie, Königstuhl 17, 69117 Heidelberg, Germany \label{inst1} \and
           Fakultät für Physik und Astronomie, Universität Heidelberg, Im Neuenheimer Feld 226, 69120 Heidelberg, Germany \label{inst2} \and
           Ludwig-Maximilians-Universität München, Universitäts-Sternwarte, Scheinerstraße 1, D-81679 München, Germany \label{inst3} \and
           Exzellenzcluster ORIGINS, Boltzmannstr. 2, D-85748 Garching, Germany \label{inst4}\\
           \email{sudarshan@mpia.de}
           }
\date{}
\abstract
{\ Protoplanetary disks observed in millimeter continuum and scattered light show a variety of substructures. While embedded planets are a common explanation, during the early planet formation phase, various physical processes in the disk could also trigger such features. One such possibility that has been previously theorized for passive disks is the thermal wave instability or the stellar irradiation instability---the flared disk may become unstable as directly illuminated regions puff up and cast shadows behind them. This would manifest as bright and dark rings, and a staircase-like structure in the disk optical surface.}
% context heading (optional) }
{\ We provide a realistic radiation hydrodynamic model to test the limits of the thermal wave instability in starlight-heated protoplanetary disks. We make quantitative comparisons to existing results in literature from simpler linear theory and 1D models to moment transfer methods, elucidating the importance of correct numerical treatment for this problem.}
{\ We carry out global axisymmetric 2D hydrostatic and dynamic simulations including radiation transport with frequency-dependent ray-traced irradiation and flux-limited diffusion (FLD). We vary dust-to-gas ratios and surface densities. We also highlight the role of small grains and dust settling with the first radiation hydrostatic dust models to study starlight-driven shadowing.}
{\ We found that starlight-driven shadows are most prominent in optically thick, slow cooling disks, shown by our models with high surface densities and dust-to-gas ratios of sub-micron grains $\edg=0.01$. We recover that thermal waves form and propagate inwards in the hydrostatic limit. In contrast, our hydrodynamic models show bumps and shadows within $30$\,au that converge to a quasi-steady state on several radiative diffusion timescales---indicating a long-lived staircase structure. We find that existing thermal pressure bumps could produce and enhance this effect, forming secondary structure due to starlight-driven shadowing downstream.
Hydrostatic models with self-consistent dust settling instead show a superheated dust irradiation absorption surface with a radially smooth temperature profile without staircases.}
{ \ We conclude that we can recover thermally induced flared-staircase structures in radiation hydrodynamic simulations of irradiated protoplanetary disks using the flux-limited diffusion method. We find that the shadowing effect is sensitive to the dust content in the disk. We highlight the importance of modeling dust dynamics consistently to explain starlight-driven shadows.}

\keywords{
          accretion disks --
          hydrodynamics --  
          radiative transfer --
          methods: numerical
          }

\maketitle
\section{Introduction} 
\label{sec:intro}
Disk substructures are ubiquitous in modern observations of planet-forming disks across different wavelengths. These include millimeter thermal emission from ground-based instruments like ALMA \citep{2018Andrews}, scattered light observations with VLT-SPHERE \citep{2018Avenhaus,2023Benisty}, and more recently, studies with JWST \citep{2025Tazaki}. Planets in the disk could carve out gaps \citep{2018Zhang} that explain substructure, of which the planetary gaps in the PDS 70 and WISPIT 2 systems are direct observational constraints \citep{2018Keppler,2019Haffert, 2025VanCap, 2025Close}. Many physical processes in the disk could also trigger substructures \citep{2023Bae}, such as zonal flows due to the magneto-rotational instability \citep[MRI,][]{2011Uribe, 2015Flock}, spirals launched by gravitational instability \citep[GI,][]{2016Kratter}, gaps and vortices resulting from the vertical shear instability \citep[VSI, ][]{2020Flock}, and spirals and rings as a consequence of misaligned inner disks \citep{2024Zhang,2025bZiampras} among others. Mechanisms such as these which could invert the local pressure gradient, radially pile up dust in the disk, acting as dust traps \citep{1972Whipple,2004Paardekooper,2026Ormel}. 

The thermal wave instability (or TWI, also known as the irradiation instability or the self-shadowing instability) is one of the mechanisms proposed to explain rings and gaps in disks \citep{2000Dullemond, 2008WatanabeLin,2012Siebenmorgen,2021Wu,2021Ueda} due to induced pressure variations and subsequent dust concentrations. Historically, this instability was first studied as a consequence of unstable solutions \citep{2000Dullemond} to the flared disk model proposed to account for the observed infrared (IR) excess in spectral energy distributions (SEDs) of passive disks around T~Tauri stars \citep{1987Kenyon,1997ChiangGoldreich,1999DAllesio}. The mechanism brings up an interesting possibility that substructures can form at the disk surface without invoking any additional ingredients, solely through starlight-driven shadowing due to variations of the irradiation angle onto the surface of a flared disk. Hints of ripples in the flared-disk scattered light observations of IM Lup with VLT-SPHERE \citep{2018Avenhaus} are particularly exciting for this scenario.

The mechanism behind this starlight-driven shadowing is outlined as follows \citep[see also Fig.~1 in][for an illustration]{2021Ueda}: in vertical hydrostatic equilibrium, the balance between gravity and pressure--- set by stellar irradiation heating-- yields a passive (irradiated) disk with flared geometry. Thus, outer regions absorb substantially more stellar light than in an internally-heated, conical disk. For optically thick disks, scale height perturbations could trigger thermal waves that travel inwards, puffing up regions of the disk and casting shadows behind them. This shows up as bright and dark bands or shadows in temperature, or staircases in the optical surface. One can refer to them as "unstable" shadows to distinguish them from the "stable" shadow cast by the hot inner rim in transition disks \citep{2001Dullemond} or due to an opacity transition at the snow line \citep{2015Bailli}.

Several methods have been used to study the starlight-driven shadowing problem in literature. Time-dependent linear theory \citep{2000Dullemond,2021Wu}, simplified 1.5D treatment \citep[i.e., mimicking vertical information with oblique radiation transfer,][]{2008WatanabeLin, 2021Ueda, 2022Okuzumi, 2022Pavlyuchenkov}, and 2.5D (i.e., 2D \{$R,z$\} axisymmetric) Monte Carlo radiative transfer models \citep{2012Siebenmorgen,2021Wu}. These aforementioned models employed either a hydrostatic or a vertically isothermal approximation to the disk (i.e., temperature fluctuations at the atmosphere propagate to the midplane instantly). In the framework of linear analysis in the optically thick disk approximation \citep{1997ChiangGoldreich,1999DAllesio, 2000Dullemond}, models predict that thermal waves form and propagate inwards. \citet{2008WatanabeLin} found quasi-periodic waves moving inwards that decayed when they reached $1$\,au. Their results also indicated that the disk could reach approximate steady states with bumps and shadows as the mass accretion rate $\dot{M}$ increases. Time-dependent analysis by \citet{2021Ueda} with a 1.5D energy equation however were focused on the models with inwardly propagating waves and found staircase structure radially outwards up to $\sim$100\,au for high dust-to-gas mass ratios. Simulations by \citet{2021Wu} with Monte-Carlo models and a global two-layer model by \citet{2022Okuzumi} that accounted for radial reprocessing of starlight within the disk, also showed inwardly propagating waves. 

However, given the coupling of radial wave propagation and vertical radiative transfer, a full radiation hydrodynamic treatment is necessary in understanding the existence and long-term implications of such an instability in realistic disks. There are several ways one can include radiation transport in hydrodynamic simulations. Those include moment methods like flux-limited diffusion \citep[FLD][]{1981LevermorePromaning} in the gray (i.e., frequency-integrated) approach \citep{1989Kley, 2013Kolb}, gray FLD coupled with frequency-dependent ray-traced irradiation \citep{2010Kuiper, 2013Flock}, frequency-dependent multi-group FLD \citep{2024Robinson}, M1 moment transfer method \citep{2021Fuksman}, a three-temperature M1 scheme \citep{2023Muley} and more complex discrete ordinate methods \citep{2021Jiang, 2022Jiang}. 

There have been some efforts in modeling the thermal wave instability with 2.5D radiation hydrodynamic simulations in recent literature. \citet{2022Pavlyuchenkov2D} used a diffusion solver for radiation transport and \citet{2022FuksmanKlahr} used M1 moment transfer coupled with frequency-dependent irradiation. While these works reported inwardly propagating thermal waves and shadows in the hydrostatic models, both studies concluded that the instability is damped in dynamical simulations. \citet{2022FuksmanKlahr} attributed this to the inefficiency in transporting thermal perturbations from the atmosphere to the midplane in the disk, as radial advection is more dominant. \citet{2024Kutra} used a simpler "radiative forcing" method which enforced the vertically isothermal approximation and found staircase structures that did not propagate inwards but settled in the disk for several orbital timescales. \referee{Recently, \citet{2025Mori} also reported thermal wave instability-like behavior in their global non-ideal magneto-hydrodynamic models, particularly in the one including the Hall effect and a magnetic field with anti-aligned polarity}.

We aim to provide limits to the thermal wave instability with a realistic radiation hydrodynamic model with frequency-dependent irradiation and cooling via FLD. We relax the vertically isothermal assumption of past models, and model advection and thermodynamics in 2.5D self-consistently. While \citet{2024Kutra} also include mass advection in 2.5D  with \texttt{ATHENA++}, our models have a more accurate treatment of radiation transfer. The ray-traced irradiation + FLD approach also models the transition from the optically thick to thin regime better  \citep{2013Kuiper} than \citet{2022Pavlyuchenkov2D}, which is more suited for optically thick disks. We also explore a higher resolution compared to them, and compare our models with the M1 moment transfer method \citep{2022FuksmanKlahr}. Finally, we extend the gas-only disk simulations to include dust settling in the hydrostatic approach, providing the first results on how dust impacts starlight-driven shadows and the thermal wave instability.
\begin{figure}
	\centering
	\includegraphics[width=\columnwidth]{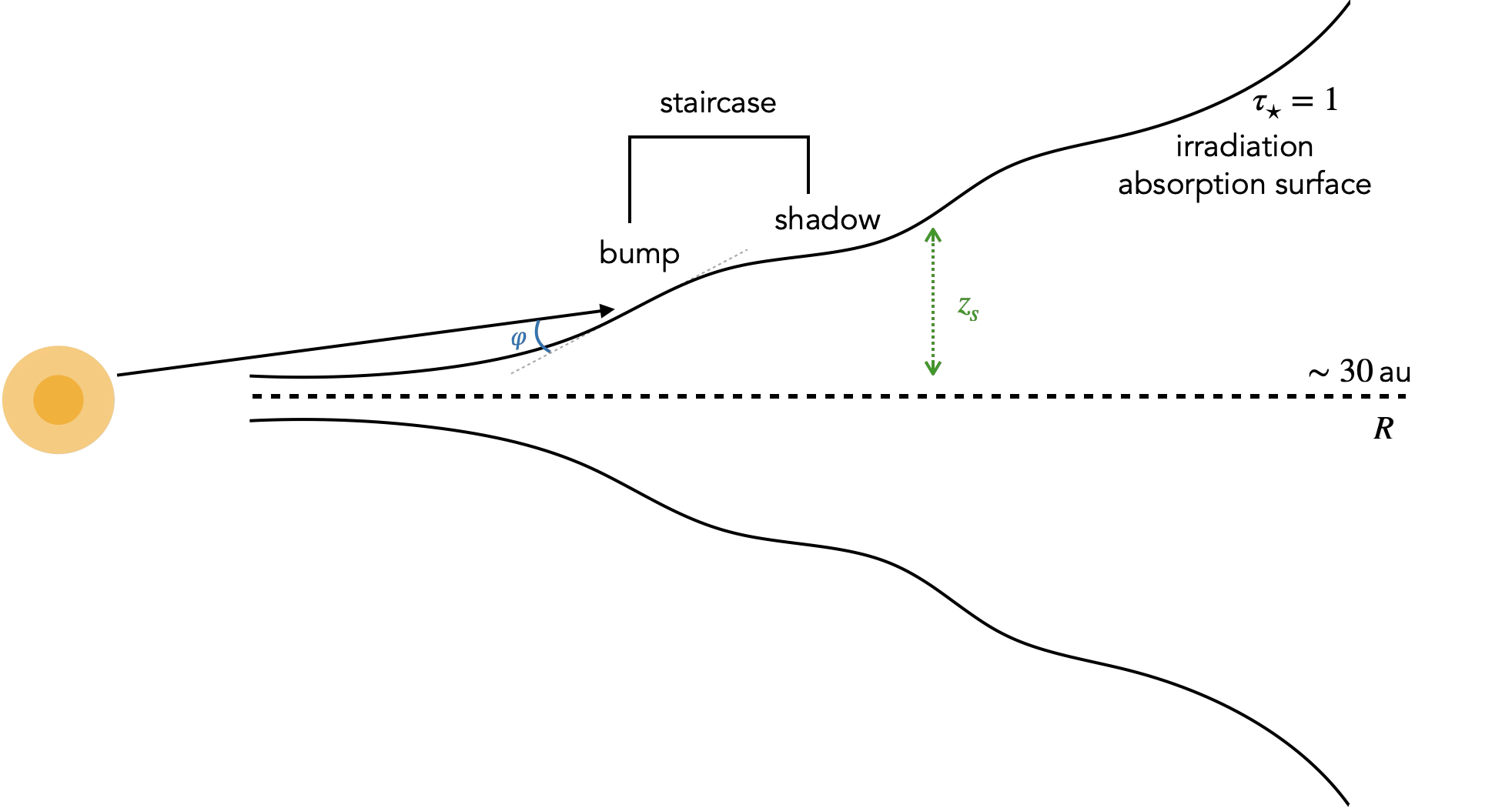}
	\caption{Schematic to illustrate the different terms: "bump", "shadow" and "staircase" defined in the introduction. The star is denoted by the orange blob, and the stellar rays hit the irradiation absorption surface $\tau_\star = 1$, at height $z_s = z~(\tau_\star=1)$ from the midplane, and a flaring angle $\varphi$.}
	\label{fig:sketchii}
\end{figure}

For clarity throughout this article, we adopt the following definitions for the words "staircase", "bump", and "shadow": a staircase is a step-like modulation in the irradiation absorption surface as a function of radius, consisting of a heated, puffed-up front (a bump) and a colder backside (a shadow). In this context, we note that a bump refers to the locally elevated, irradiated region, rather than the more traditional temperature/pressure gradient inversion commonly associated with the term. A shadow, in turn, corresponds to the dip in the irradiation flux downstream, which we trace through the midplane temperature decrease of at least 5\% relative to the background power-law, this is clearly illustrated by the sketch in Fig.~\ref{fig:sketchii}. In Sect.~\ref{sec:theory} we outline the physics used in our model, the model parameters, and the numerical methods used. In Sects.~\ref{sec:results-fiducial}--\ref{sec:results-dust} we show the results from our hydrostatic, dynamical, and dust models. We then discuss the implications of our results in the context of previous work in Sect.~\ref{sec:discussion} and summarize our findings in Sect.~\ref{sec:summary}.

\section{Physics, model setup, \editor{and numerics}}
\label{sec:theory}
In this section, we outline the physics used in our model. We start with the gas dynamics and then describe the radiation method,
whose self-consistent treatment is crucial for capturing
how the disk responds to thermal perturbations. 

\subsection{Radiation hydrodynamic equations}
\label{sec:rad-hydro}

We write down the conservative equations for gas dynamics and radiation transport in the gray flux-limited diffusion approximation following \citet{2013Flock}. For gas density $\rho$, velocity $\bm{u}$, total energy density $\Etot$, radiation energy density $\Erad$ and radiation flux $\bmFrad$, they read as

\begin{subequations}
	\label{eq:conservative}
	\begin{align}\label{eq:conservative-1}
        \DP{\rho}{t} + \nabla&\cdot{(\rho\bm{u})} = 0 ,
	\end{align}
	\begin{equation}\label{eq:conservative-2}
    \begin{aligned}
	\DP{(\rho\bm{u})}{t} + \nabla&\cdot(\rho\bm{u}\otimes\bm{u} + P \bm{\mathrm{I}}) 
         = - \rho\nabla\phistar,
    \end{aligned}
	\end{equation}
	\begin{equation}\label{eq:conservative-3}
    \begin{aligned}
	\DP{\Etot}{t} + \nabla&\cdot[(\Etot + P)\,\bm{u}] =
       - \rho\bm{u}\cdot\nabla\phistar\\ &
        - \kp\rho c \left(\arad T^4 - \Erad\right) - \nabla\cdot\Fstar
    \end{aligned}
	\end{equation}
	\begin{equation}\label{eq:conservative-4}
    \begin{aligned}
        \DP{\Erad}{t} + \nabla&\cdot\bmFrad = \kp\rho c \left(\arad T^4 - \Erad\right), \\
        &\bmFrad = -\frac{\lambda c}{\kr\rho} \nabla\Erad.
	\end{aligned}
    \end{equation}
\end{subequations}

The gravitational potential of the star is given by $\phistar = - \G\Mstar/r$ where $\G$ is the gravitational constant, $\Mstar$ is the mass of the central star, and $r=\sqrt{R^2+z^2}$ is the spherical radius distance to the star. The total energy is a function of kinetic energy and the specific internal energy $e$ with $\Etot = \rho e + \frac{1}{2} \rho\bm{u}^2$. The gas pressure is denoted by $P$, and $\bm{\mathrm{I}}$ is the unit matrix. The temperature-dependent Planck and Rosseland mean opacities are $\kp$ and $\kr$ respectively, $\arad$ is the radiation constant, and $c$ is the speed of light. The set of equations is complete with a closure relation, that is, the ideal gas equation $P = (\gamma - 1) \rho e$ with the adiabatic index chosen as $\gamma = 1.42$ \citep{1978Decampli} and the mean molecular weight $\mu = 2.35$. The gas temperature $T$ is given by $e = \rho c_\text{V} T$, with $c_\text{V} = R_\text{gas} / \mu (\gamma - 1)$ being the specific heat capacity at constant volume, and $R_\text{gas}$ the gas constant. The isothermal sound speed is defined as $\cs = \sqrt{P/\rho}$ and the gas pressure scale height is given by $H = \cs/\OmegaK$ where $\OmegaK = \sqrt{G\Mstar/R^3}$ is the Keplerian angular velocity.

The two coupled equations for radiative transfer are given by
\vspace{-1em}
\begin{subequations}
    \label{eq:fld-step}
	\begin{align}
        \label{eq:fld-step-1}
	    \DP{e}{t} &= - \kp\rho c \left(\arad T^4 - \Erad\right) - \nabla\cdot\Fstar,
	\end{align}\vspace{-2em}
	\begin{align}
        \label{eq:fld-step-2}
        \DP{\Erad}{t} &+ \nabla\cdot\bmFrad = \kp\rho c \left(\arad T^4 - \Erad\right).
	\end{align}
\end{subequations}
We write the radiation flux above in terms of the radiation energy density with a flux limiter $\lambda$  which accounts for the optically thick ($\lambda \rightarrow 1/3$) and thin regimes \citep[$\Frad \rightarrow c \Erad$, see also][]{2013Kolb}. We use the flux limiter of \citet{1981LevermorePromaning} which is given by $\lambda = \frac{2 + X}{6 + 3X+ X^2}$ with $X = \frac{|\nabla\Erad|}{\kr\rho\Erad}$.

\subsection{Frequency–dependent irradiation}
\label{sec:freq-irrad}
The transition between the optically thick midplane of the disk to the optically thin surface is better captured by including ray-traced irradiation from the star at different frequencies \citep{2010Kuiper}.
The frequency–dependent irradiation flux $\Fstar$ from the central star is written in terms of the black body spectrum $B_\nu (\Tstar)$,

\begin{equation}\label{eq:irrad-flux}
\Fstar (r) = \Int_{\nu} \pi B_\nu (\Tstar) 
\left(\frac{\Rstar}{r}\right)^2 e^{-\tau_\nu(r)} \,d\nu \,,
\end{equation}
  
with the optical depth $\tau_\nu$ defined as an integral over the solid angle $\Omega$ and frequency $\nu$ (or wavelength $\lambda=c/\nu$),
\begin{equation}\label{eq:irrad-tau}
    \tau_\nu(r) = \Int_{\Rstar}^r \knu \rho \,dr = \tau_{\nu,0}\, (\rbeg) + \Int_{\rbeg}^r\knu \rho \,dr .
\end{equation}

\begin{figure}
	\centering
	\includegraphics[width=\columnwidth]{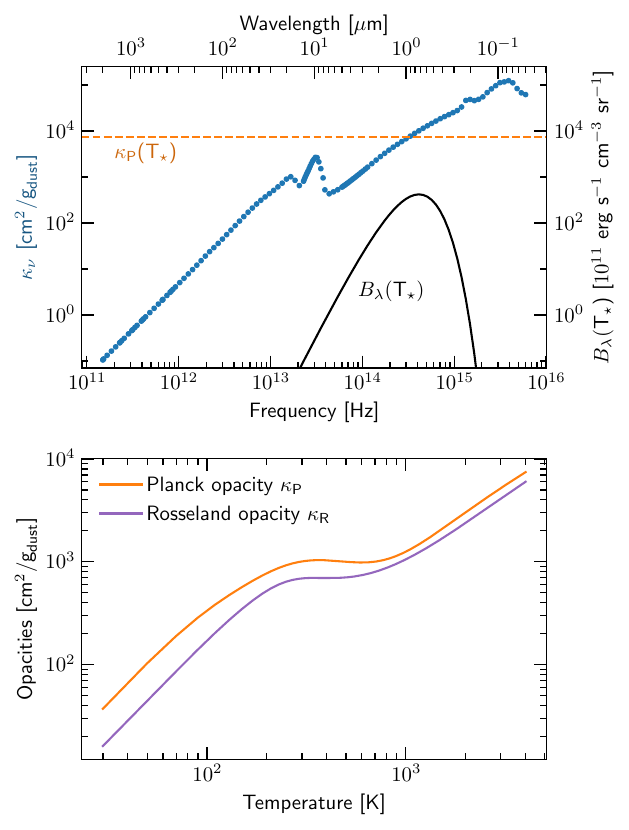}
	\caption{Top panel: Frequency-dependent dust absorption opacities (per gram of dust, $\text{g}_\text{dust}$) from tabulated values \citep{2020AKriegerWolf,2022KriegerWolf}. The blue points denote 132 frequencies logarithmically sampled in the range $\nu \in [1.5 \times 10^{11}, 6 \times 10^{15}]$~Hz for the computation of irradiation flux in our models. The black line indicates the black-body function and the orange line is the Planck opacity for $\Tstar = 4000~$K \citep[figure similar to][]{2010Kuiper}. Bottom panel: Temperature–dependent Planck and Rosseland mean opacities used for radiation transfer. The local maximum ($\approx$ 250\,K) and the subsequent dip in the mean opacities is due to the $10\,\mu$m silicate feature.}
	\label{fig:opacfig}
\end{figure}

 The optical depth is a function of the frequency-dependent absorption opacity $\knu$ and the density $\rho$. Following \citet{2022FuksmanKlahr}, we use tabulated values for opacities provided by the dust from \citet{2020AKriegerWolf,2022KriegerWolf}. We plot $\knu$, $\kp$, and $\kr$, along with the blackbody function $B_\lambda (\Tstar) = \nicefrac{c}{\lambda^2} B_\nu (\Tstar)$ for $\Tstar = 4000$\,K in Fig.~\ref{fig:opacfig}. We do not model the hot inner rim of the disk that otherwise casts a shadow, and instead compute an effective optical depth $\tau_{\nu, 0}$ from the central star out to the edge $\rbeg$ of the disk region of particular interest for our problem, given by
 \begin{equation}
 \label{eq:tau0}
     \tau_{\nu, 0} (\rbeg)= \knu \rho_{\rbeg} (\rbeg - \Rstar)
 \end{equation}
 following a similar prescription from \citet{2013Flock}. 
A more detailed discussion on this initial optical depth profile and how it aligns with our current understanding of the inner disk structure can be found in Sect.~\ref{sec:inner-disk}. We then calculate the irradiation flux at each cell center by ray-tracing for different frequency bins. 
 
\subsection{Model parameters}
\label{sec:model-params}
For our reference run \texttt{MFID} we model a T~Tauri central star, similar to previous models in the literature \citep[e.g.,][]{2008WatanabeLin,2021Ueda, 2022FuksmanKlahr}, with a mass $\Mstar = 1\,\Msun$, radius $\Rstar = 2.6\,\Rsun$, effective temperature $\Tstar = 4000\,$K, and luminosity $\Lstar = 1.56\,\mathrm{L}_\odot$. We implement a surface density power law of $\Sigma (R) = \Sigma_0\,(R/1\,\text{au})^{-1}$, with $R$ being the cylindrical radius. We use $\Sigma_0 = 200\,\mathrm{g\,cm}^{-2}$ for the \texttt{MFID} model, consistent with dust-mass based measurements at $1\,\mathrm{au}$ \citep{2017VanBoekel}. To explore the effects of higher disk masses, we also consider a model with $\Sigma_0 = 2000\,\mathrm{g\,cm}^{-2}$ (\texttt{MSIG2000}), which is more representative of classical minimum-mass solar nebula models \citep[MMSN,][]{1977Weidenschilling,1981Hayashi}. While dust mass estimates support the lower surface densities, we explore both models as gas mass estimates still remain poorly constrained. The higher surface density profile matches the one computed using a viscous disk model with a mass accretion rate $\dot{M} = 10^{-8}\,\Msun\,\text{yr}^{-1}$ and viscosity parameter $\alpha = 10^{-3}$ following \citet{1973ShakuraSunyaev}. This is a commonly used initial condition in disk theoretical studies \citep[including the thermal wave instability, e.g.,][]{2008WatanabeLin, 2019Flock,2021Steiner}. However, our models ignore viscous heating \citep[including it would yield a lower $\Sigma$ in the inner regions, see Fig.~3 of ][]{2020Ziampras}, which results in a disk that is marginally more massive in the innermost regions ($\lesssim 5\,\text{au}$). 

The initial guess for the temperature is determined by the optically thin solution for a passively irradiated disk \citep{1997ChiangGoldreich}. We also sample dust-to-gas mass ratios $\edg =\{0.01, 0.001\}$ (models \texttt{MFID} and \texttt{MD0.001}, respectively). A table with the different setups corresponding to the model parameters is given in Table~\ref{tab:modelparams}. This excludes the hydrostatic dust models for which the  parameters are described in detail in Sect.~\ref{sec:dust-scale}. Our models are inviscid and do not account for the effect of accretion heating.
 \begin{table}
        \caption{Parameters for the different hydrostatic and dynamic runs.}
        \centering
        \begin{tabular}{ l  c c c c c}
          \hline
          Model & $\edg$ & $\Lstar\,(\mathrm{L}_\odot)$ & $\Tstar$\,(K) & $\Rstar\,(\mathrm{R}_\odot)$ & $\Sigma_0\,(\nicefrac{\text{g}}{\text{cm}^2})$ \\
          \hline 
          \rule{0pt}{2.5ex}\texttt{MFID} & $0.01$ & $1.56$ & $4000$ & $2.6$ & $200$ \\ \texttt{MD0.001} &  $0.001$ & $1.56$ & $4000$ & $2.6$ & $200$ \\ 
          \texttt{MSIG2000}&  $0.01$ & $1.56$ & $4000$ & $2.6$ & $2000$ \\
          \texttt{ML1}\tablefootmark{*}&  $0.01$ & $1$ & $4000$ & $2.086$ & $200$ \\
          \hline \\
        \end{tabular}
        \tablefoot{Each model is denoted by \texttt{M} followed by a label indicating the parameter altered relative to the fiducial run \texttt{MFID}. \\
        \tablefoottext{*}{\texttt{ML1} is an additional model we set up that matches \citet{2022FuksmanKlahr} with a stellar parameters corresponding to $\Lstar = 1\,\mathrm{L}_\odot$, that we only discuss during the comparison in Appendix \ref{sec:fld-m1}.}}
        \label{tab:modelparams}
\end{table}

\subsection{\editor{Numerics}}
\label{sec:numerics}
We solve the radiation hydrodynamic equations outlined in Eqs.~\eqref{eq:conservative} and \eqref{eq:fld-step} using the astrophysical finite–volume fluid code \texttt{PLUTO} \citep{2007Mignone}. We use the flux–limited diffusion module outlined in \citet{2013Flock} which uses a Jacobi preconditioned BiCGSTAB solver for the radiation matrix. We perform 2.5D axisymmetric simulations of gas assuming a constant dust-to-gas mass ratio $\edg$ in spherical coordinates ($r, \theta$) with a logarithmic grid in $r \in [0.41, 95]$ au and a uniform grid in $\theta \in [\nicefrac{\pi}{2} - 0.32, \nicefrac{\pi}{2} + 0.32]$. We use a grid of $N_r \times N_\theta = 1024 \times 128$ cells, which corresponds to a resolution of $\approx6$ cps (cells per scale height) in $r$ and $\approx7~$cps in $\theta$. We use a second-order Runge Kutta (RK2) time integrator, a linear reconstruction scheme, the HLLc Riemann solver \citep{1994Toro} and a CFL (Courant--Friedrichs--Lewy) value of 0.4.

 We implement strict outflow conditions for the radial velocity $u_r$ at the radial inner boundary ($R < 0.5$~au), and use power-law extrapolation for the azimuthal velocity $u_\phi$. The rest of the quantities are set to zero gradient. We also damp the density and velocities at the inner boundary using the \citet{2006DeValBorro} prescription over a damping timescale of $0.01$ inner orbits at $R_0$. For the radiation boundaries, we restrict the radiation energy density at the polar boundaries to a small value $\Erad = a_R T_0^4$, where $T_0 =  370 \,\edg\,(R/\text{au})^{-0.5}$~K which is about 7\% of the optically thin temperature profile for the model \texttt{MFID} \citep{1997ChiangGoldreich}. Following the prescription in \citet{2016Flock} (refer Eq.~(19) in the paper), we implement a modified gravitational potential in the inner radial boundary to improve numerical stability in the dynamical simulations.  In very optical thin regions of the disk we modify the Rosseland mean opacity $\kappa_R$ so that $\knu\rho_{\rbeg} \Delta r > 10^{-4}$ is fulfilled. This helps the convergence of the matrix solver without affecting the temperature significantly. Furthermore, we ensure that the radially innermost cell remains optically thin, preventing non-physical heating by the irradiation flux \citep{2013Flock}.

\subsection{Iterative solver for hydrostatic models}
\label{sec:iterative-solver}

Using the iterative method described in \citet{2013Flock} to compute solutions for irradiated disks in radiative hydrostatic equilibrium, we first run hydrostatic models with the parameters described above. Given an initial density profile $\rho (r, \theta)$, this solver iterates between each FLD step given by Eq.~\eqref{eq:fld-step} over a typical radiative diffusion timescale of $\sim$$10^6$\,s (for typical solar parameters, see also Eq.~\eqref{eq:cooling-timescale}). The extended runtimes with successive diffusion steps reaching up to $10^8$\,s in later iterations ensure thermal equilibrium, with temperatures reaching a converged state. Once the radiation solver converges, new densities (and azimuthal velocities) are computed by integrating the hydrostatic equations,
\begin{subequations}
        \label{eq:hydrostatic}
        \begin{align}\label{eq:hydrostatic-1}
        \DP{P}{r} &= - \rho \DP{\Phi_\star}{r} + \frac{\rho u_\phi^2}{r},
        \end{align}
        \begin{align}\label{eq:hydrostatic-2}
        \frac{1}{r}\DP{P}{\theta} &= \text{cot}\theta \, \frac{\rho u_\phi^2}{r},
        \end{align}             
\end{subequations}
using a second-order Runge--Kutta integrator. The iterative process is continued until a converged density state is reached. The computed profiles are then used as initial conditions for the hydrodynamic simulations. While disk hydrostatics are more correctly treated in cylindrical coordinates \citep[see, e.g., codes like \texttt{cuDisc},][]{2024Robinson}, our approach works well and has been robustly tested and used for various problems \citep[for validation tests, see Appendix~B of][]{2013Flock}. \\

Previous models of the thermal wave instability suggest that unstable shadows form and propagate inwards in such a hydrostatic iterative treatment \citep{2021Ueda,2022Okuzumi}. Our models further include vertical and radial radiative transfer, relaxing the commonly used isothermal approximation. We start by describing these hydrostatic models, followed by the hydrodynamic simulations in the following sections.

\section{Results I: Fiducial model \texttt{MFID}} 
\label{sec:results-fiducial}

We first outline results from our fiducial model \texttt{MFID}, for which the model parameters are listed in Table~\ref{tab:modelparams}. This corresponds to a high dust-to-gas ratio case of $\edg = 0.01$. Such a value represents the early times of disk evolution, where most of the dust material is still in the form of small ($\leq 10$\,$\mu$m) grain sizes. For reference, a total dust-to-gas mass ratio $\epsilon_\text{tot} = 0.01$ for all sizes and MRN dust distribution \citep{1977Mathis}, and a dust-to-gas ratio in small grains $\edg$ of $0.01$, $0.001$ and $0.0001$ would yield maximum grain sizes $a_\text{max}$ of $0.25\,\mu$m, $19\,\mu$m and $1.8$\,mm, respectively \citep[see also][]{2024Fuksman}. At later stages, dust grains grow,  settle to the midplane, and small grains become depleted in the disk \citep{2010Birnstiel,2014Testi}. While $\edg = 0.01$ may be on the higher end for a T~Tauri disk, we adopt this value as our fiducial case in order to optimize the conditions under which staircase structures are most likely to develop.

\subsection{Hydrostatic model}
\label{sec:hydrostatic-high-d2g}
\begin{figure}[h]
	\centering
	\includegraphics[width=\columnwidth]{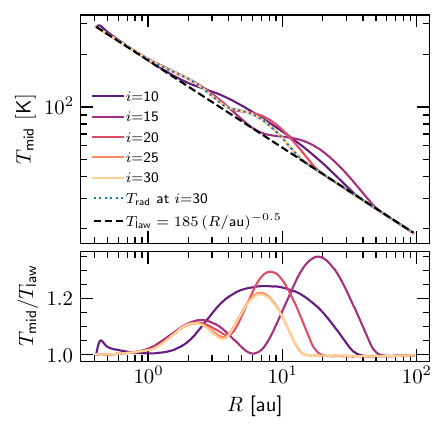}
	\caption{Top: Time evolution of the temperature profile at the disk midplane for iterations $i=[10,15,20,25,30]$. The profiles indicate that thermal waves form and travel inwards with time. There are two bumps/shadows at the end of the hydrostatic run for model \texttt{MFID}. The blue dotted line is the radiation temperature at $i=30$ and the black dashed line indicates the background power law fit for the midplane $T_{\text{law}} = 185\,(R/\text{au})^{-0.5}$. Bottom: Deviation of the temperature profile from the background power law.}
	\label{fig:timeevolstat}
\end{figure}

\begin{figure*}[ht]
        \centering
        \includegraphics[width=\linewidth]{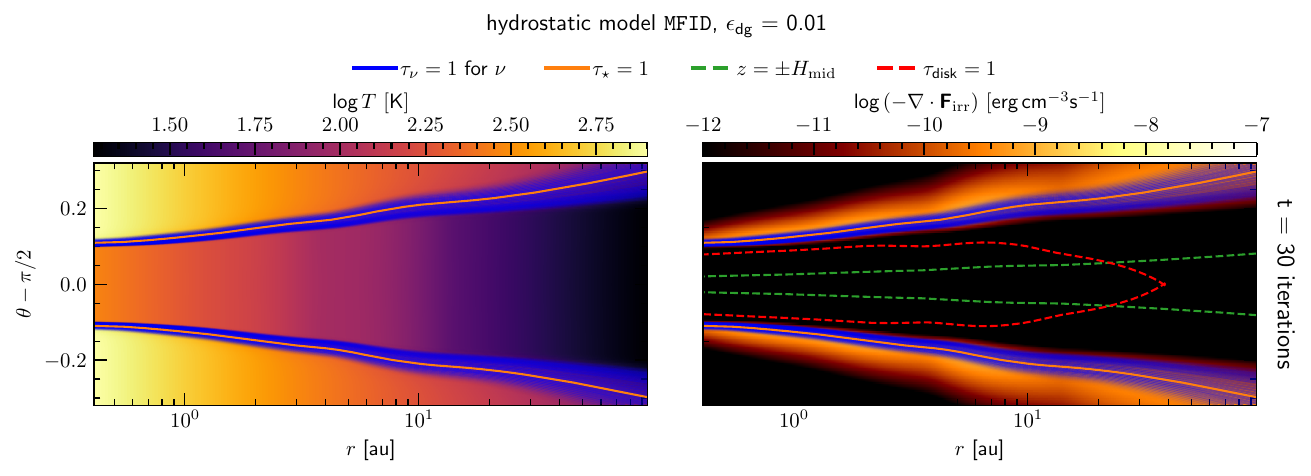}
        \caption{2D profiles of temperature and the irradiation heating term for the hydrostatic model at $i = 30$ iterations. The orange line refers to the $\tau_\star$ = 1 profile for the stellar radiation using the Planck mean opacity. Blue lines refer to $\tau_\nu = 1$ profiles for for different frequencies $\nu$ (same as in Fig.~\ref{fig:opacfig}), with the line opacity weighted by $B_\nu$ at that frequency. \citep[see also Fig.~7 in ][]{2022FuksmanKlahr} . The red line indicates the $\tau_\text{disk} = 1$ profile for the disk's radiation using $\kappa_P (T)$. The green lines indicate the midplane gas scale height of the disk.}
        \label{fig:tempfluxstat}
\end{figure*}

We see the emergence of alternating bright and dark regions in the disk with thermal waves forming and propagating inwards, consistent with the expectations from hydrostatic models in the literature \citep{2008WatanabeLin, 2021Ueda,2022Pavlyuchenkov}. This is clearly depicted by the midplane temperature profiles plotted for different iteration times in Fig.~\ref{fig:timeevolstat} and also the 2D snapshot of temperature and irradiation heating rate $-\nabla\cdot F_\star$ at the end of the hydrostatic run in Fig.~\ref{fig:tempfluxstat}.
 
From Fig.~\ref{fig:timeevolstat}, we see that as the disk tries to find a hydrostatic solution, a thermal wave forms and travels inwards, and two form after $i=15$ iterations that also propagate inwards. The perturbations in temperature deviate from the background power law by up to 30\%.
\begin{figure*}[ht]
        \centering
        \includegraphics[width=\linewidth]{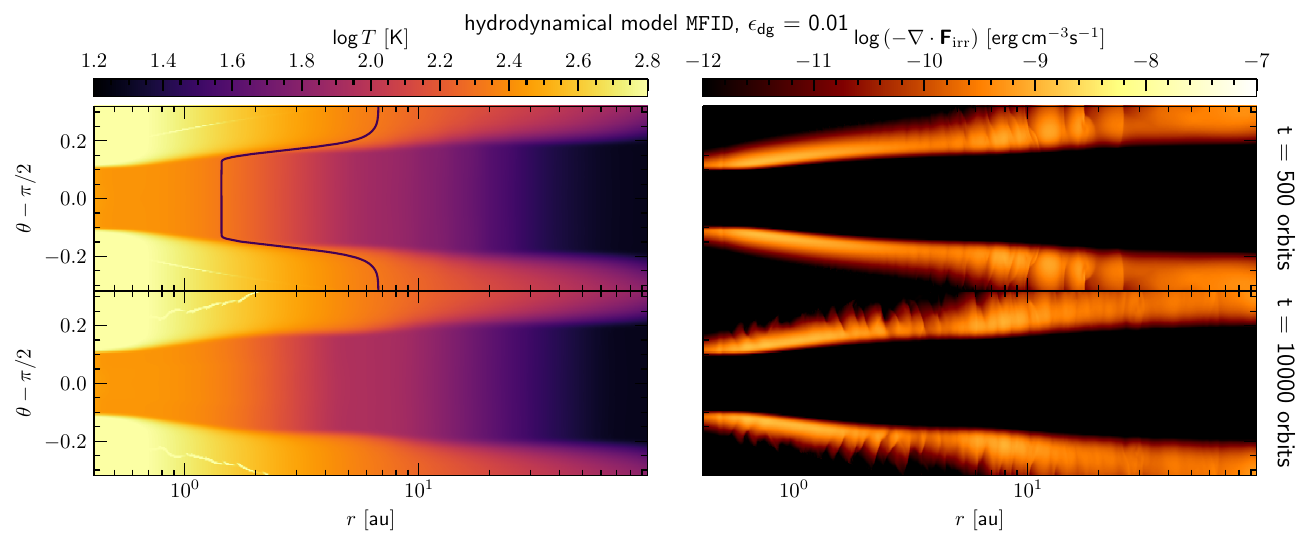}
        \caption{2D profiles of temperature and the irradiation heating term for the hydrodynamic model at $t = 500$ orbits (top) and $t = 10000$ orbits (bottom). We see that the inner staircase (corresponding to the hydrostatic model in Fig.~\ref{fig:timeevolstat}) has grown and slightly shifted inwards with the outer staircase readjusting as a consequence of the first. The black solid line denotes the temperature corresponding to the opacity transition in Fig.~\ref{fig:opacfig}. We see that these staircases remain in the disk for thousands of orbits spanning several radiative diffusion timescales.}
        \label{fig:dyn-0.01}
\end{figure*}
\begin{figure}[h]
	\centering
	\includegraphics[width=\columnwidth]{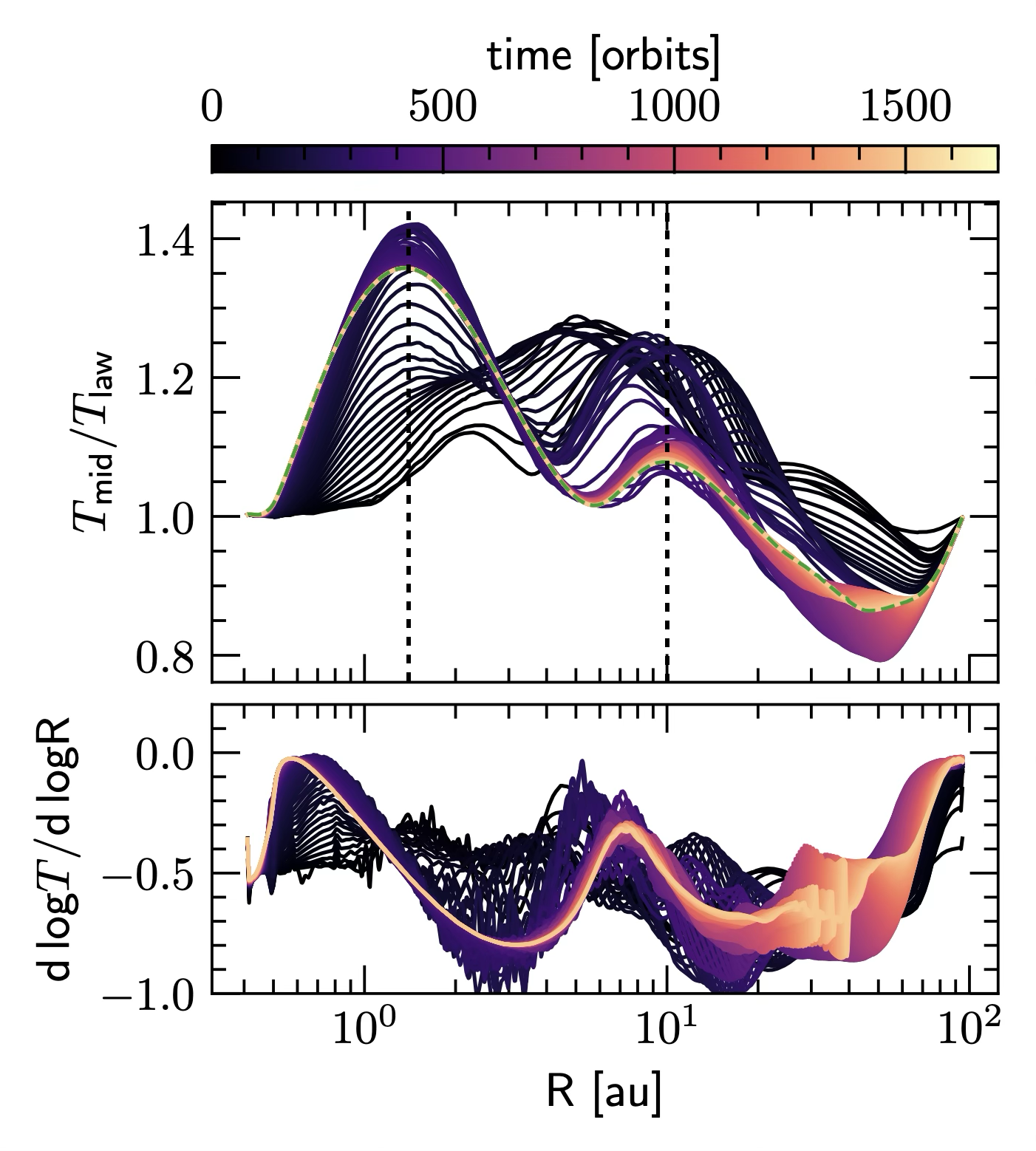}
	\caption{Top: Time evolution of the temperature profile at the disk midplane against the background power law. We see here that the bumps are not moving inwards chaotically, and while we lose the outer bump, the inner two stabilize at about $t = 400$ orbits. The first bump grows in amplitude, and as a result casts a steeper shadow behind, moving the second shadow further than where it started. The dashed lines (the first corresponds to the opacity transition) indicate the final locations of the temperature bumps. Bottom: Time evolution of the radial temperature gradient. In shadowed regions the profile gets less steep, but overall the "bumps" are not strong enough to induce a sign flip.}
	\label{fig:dyn-time-evol}
\end{figure}

We plot the 2D snapshots at the last iteration ($i = 30$) 
in  Fig.~\ref{fig:tempfluxstat}. To get a complete picture of where in the disk radiation is emitted and absorbed, we plot different optical depth surfaces, that is, $\tau_\nu = 1$ profiles corresponding to the starlight at different frequencies $\tau_\nu$ and the disk's own emission $\tau_{\text{disk}}$ defined by
\begin{equation}
        \label{eq:tauirrdisk}
            \tau_\nu = \int_{\Rstar}^{r} \kappa_\nu \rho dr, \quad \tau_\text{disk} = \int_{\theta = \pi/2}^{\theta} \kappa_\nu \rho r d\theta  .
        \end{equation}
The blue shaded regions correspond to $\tau_\nu = 1$ surfaces for different frequencies $\nu$ (spanning the range in Fig.~\ref{fig:opacfig}), and the orange line is the $\tau_{\star} = 1$ surface computed using the Planck mean opacity at the stellar temperature $\kp (\Tstar)$. These curves exhibit a staircase pattern in contrast to the typical smooth-flared disk power law solution. Nevertheless, it is important to investigate the existence of such a pattern in hydrodynamic simulations to assess their stability. This is discussed in the following section. The red line in Fig.~\ref{fig:tempfluxstat} indicates the optical depth to the disk's own emission $\tau_{\text{disk}} = 1$, indicating that the disk is optically thick up to $R \sim 50$\,au.

\subsection{Hydrodynamics}
\label{sec:dynamical-high-d2g}
With initial conditions from the previously described hydrostatic model, we introduce dynamics, allowing the system to evolve over time. Plotted in Fig.~\ref{fig:dyn-0.01} (top panel) are the profiles of the temperature and the irradiation source term at $t = 500$~inner orbits or 131.3\,yrs. We do not find inwardly propagating waves. Instead, among the bumps in temperature that we started out with in the hydrostatic case, the innermost bump grows in amplitude, and as a result of this bump, the region behind it is shadowed, giving rise to the second one. After about 400 inner orbits, we see that this staircase structure reaches a quasi-steady state that remains stable for several hundreds of orbits. This is clearer in the top panel of Fig.~\ref{fig:dyn-time-evol}, where we plot radial profiles of the midplane temperature normalized to the background power law for different times.

\subsubsection{The inner bump}
\label{sec:inner-bump}
The innermost bump has slightly grown, and shifted inwards compared to its radial location in the static models. There are two factors to consider while interpreting this first bump as a consequence of the thermal wave instability. First, its location between 1--2\,au, which corresponds to a temperature of $\approx250$\,K, coincides with the $10\,\mu$m silicate transition\footnote{This is more of a local change in the power-law dependence of the opacity on temperature due to variations in the optical constants near the silicate resonance rather than a jump due to a change in dust composition (e.g., near a snowline).} in the temperature-dependent opacities (see the bottom panel of Fig.~\ref{fig:opacfig}). This is further indicated by a black solid line in Fig.~\ref{fig:dyn-0.01}, and the first dashed line in the time evolution plot in Fig.~\ref{fig:dyn-time-evol}. Regions in the disk with opacity transitions can locally change the heating rate, and a strong enough transition could also flatten the pressure gradient causing bumps \citep{2007GaraudLin}. While this may not be enough for the formation of this inner bump, it may have an influence. Second, the inner disk is modeled with an effective optical depth profile $\tau_{\nu, 0}$, which can affect the first bump through a non-uniform heating rate. In our simulations, we do find a slight enhancement in the irradiation power inwards of $1\,$au (see Fig.~\ref{fig:tau0-comp}) with the prescription of $\tau_{\nu, 0}$. A more detailed discussion on this is included in Appendix~\ref{sec:tau0-sampling}.

\subsubsection{Secondary structure around a bump}
Regardless of the origin of the first bump, we would like to highlight that the key consequence of this inner bump is  that it casts a shadow behind, driving a second one. This second bump is a natural consequence of the first, driven by starlight-driven shadowing. This behavior is unlike that expected by the ``thermal wave instability'' discussed in \citet{2021Wu}~and~\citet{2021Ueda}, where the staircase pattern propagates inwards. Instead, the disk seems to relax to a quasi-steady state characterized by bumps and shadows formed sequentially from the inside out. Forming secondary structure around existing bumps creates interesting implications for disk studies focused on pressure traps (see Sect.~\ref{sec:snowlines} for a discussion on snowlines). 

We plot the radial temperature gradient in the bottom panel of Fig.~\ref{fig:dyn-time-evol}. The bumps/shadows have a clear imprint with modulations of the radial temperature profile. However, neither the first bump/shadow or the consequent ones in the \texttt{MFID} model produce local extrema ($\nicefrac{\text{d}T_{\text{mid}}}{\text{d}R} = 0$)---the profile remains radially monotonic, albeit with a varying slope.
\subsection{Long-term evolution of the staircase structure}
\label{sec:long-term-1}
We run the fiducial model \texttt{MFID} for a long time, the second panel of Fig.~\ref{fig:dyn-0.01} shows the 2D profiles after $t = 10000$~inner orbits or 2625.3\,yrs. We see that the two bumps/shadows remain in the disk after several thousands of years. This brings up the interesting possibility that they could correspond to long-lived substructure in the disk. Both panels showing temperature in Fig.~\ref{fig:dyn-0.01} show hot regions in the upper layers of the inner disk, owing to numerical artifacts that arise due to a combination of strong shocks at very low-density regions and some compressional heating due to the modified gravitational potential \citep[seen also in other works like][]{2022FuksmanKlahr, 2024Cecil}.

Our results help support the fact that starlight-driven staircases are dramatically affected by hydrodynamics in the disk, and that predictions from linear theory and hydrostatic models do not fully hold up.
Our results for the $\edg = 0.01$ \texttt{MFID} model are different from such predictions where inwardly propagating thermal waves were seen \citep{2021Wu,2021Ueda,2022Pavlyuchenkov}. This is also different from the dynamical models of \citet{2022FuksmanKlahr}, who found that the thermal waves completely damp away, although our setups have a different treatment of the inner disk profile. We show that in the dynamical case, while thermal waves do not propagate inwards, they grow and reach a quasi-steady state. While \citet{2024Kutra} hint towards a similar theory of staircases settling in the disk, there are clear differences in the physics we model. We make comparisons to previous works in detail in Section~\ref{sec:discussion}.

\section{Results II: Changing the dust content}
\label{sec:parameterstudy}
We explore the two models \texttt{MD0.001} and \texttt{MSIG2000} with lower dust-to-gas ratios of $\edg = 0.001$ and a higher surface density $\Sigma_0 = 2000~\text{g/cm}^2$ respectively. Both of these vary the dust content in the disk, which controls the efficiency of radiation transport and hence affects the thermal structure of the disk. The growth rate of the instability, predicted by \citet{2021Ueda}, \citet{2021Wu} depends on the height at which the irradiation is absorbed on the surface of the disk. Disks with a higher dust opacity, be it due to a higher $\edg$ or gas mass, achieve this. Previous work also verified that for lower dust content, the TWI was suppressed \citep{2021Ueda, 2022FuksmanKlahr}.

\subsection{Low dust-to-gas ratio model \texttt{MD0.001}}
\label{sec:low-d2g}

\begin{figure*}[ht]
        \centering
        \includegraphics[width=\linewidth]{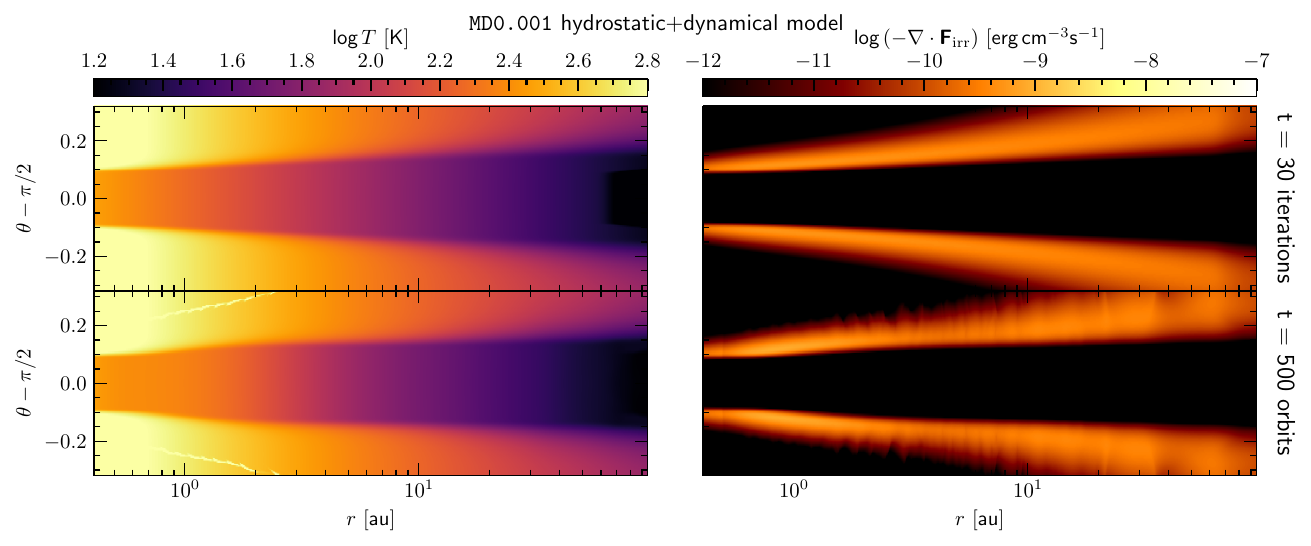}
        \caption{2D profiles of temperature and the irradiation source term for the \texttt{MD0.001} hydrostatic+dynamic model with $\edg = 0.001$. Top panel: Hydrostatic model     after $t = 30$ iterations does not show any staircase structure. Bottom panel: Hydrodynamic model after $t = 500$ orbits shows the innermost bump.}
        \label{fig:dglow}
\end{figure*}
\begin{figure*}[ht]
        \centering
        \includegraphics[width=\linewidth]{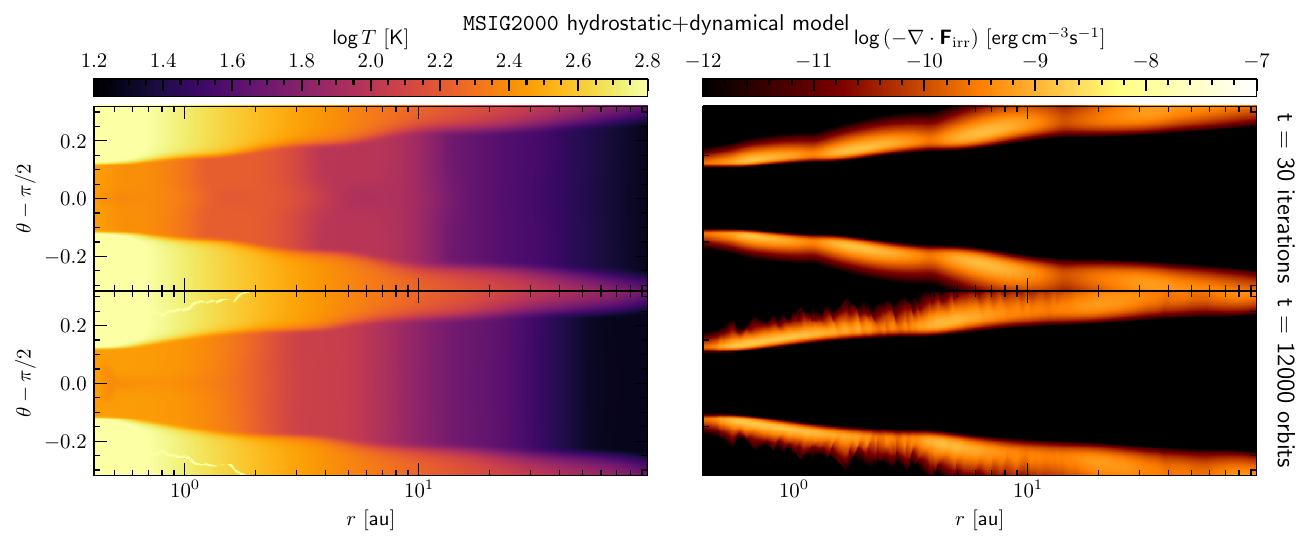}
        \caption{2D profiles of temperature and the irradiation source term for the high surface density \texttt{MSIG2000} hydrostatic+dynamic  model. The hydrostatic run shows three prominent bumps (top), hydrodynamic run shows two bumps (bottom). Midplane temperature reached equilibrium after several radiative diffusion timescales, and hence the longer run time of $t \approx 12000$ orbits.}
        \label{fig:sigma-10}
\end{figure*}
In this section, we outline our results from our hydrostatic and dynamical models for the low dust-to-gas ratio case of $\edg = 0.001$. As outlined above, these values are more typical to the dust content observed in later stages \citep[i.e., Class II protoplanetary disks][]{2021Rilinger, 2024Birnstiel}. Figure~\ref{fig:dglow} shows the temperature and the irradiation source term after 30 iterations for the hydrostatic model (top) and the hydrodynamic model (bottom) after $t = 500$ orbits. We do not see any temperature bumps/shadows forming in the hydrostatic model, but recover an inner bump in the dynamic model. 

While studies in literature point towards the TWI being suppressed for very low dust-to-gas mass ratios $\edg \lessapprox 10^{-4}$ \citep{2021Ueda}, we do not recover any secondary shadowing already for the $\edg = 0.001$ model. The combined effect of the reduced dust content and a lower radiation boundary (see $T_0$ in Sect.~\ref{sec:numerics}) could enhance the cooling efficiency of the disk in the outer parts, affecting the formation of staircases. An outline of the cooling timescales in the disk can be found in the Appendix~\ref{sec:cooling-timescales}. We speculate that the amplitude of the shadowing being smaller using the FLD method could also impact this limit. We discuss this further in Sect.~\ref{sec:amplitude}.

\subsection{High surface density model \texttt{MSIG2000}}
\label{sec:high-sigma}

\begin{figure}[h]
	\centering
	\includegraphics[width=\columnwidth]{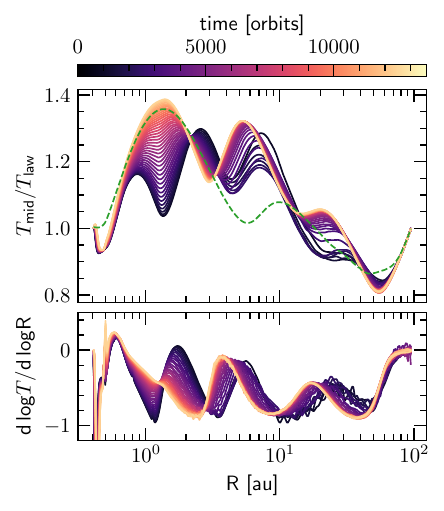}
	\caption{Top: Long term time evolution of the midplane temperature deviation from the background power law for the dynamical model \texttt{MSIG2000}. We see that among the three bumps we begin with, the innermost grows in amplitude, while the second and third move inwards. After about $t = 700$ orbits, we see that the first and the second bump merge together. One can also spot an additional bump at $\approx 20$~au. The green dashed line corresponds to the final state of the \texttt{MFID} model in Fig.~\ref{fig:dyn-time-evol}, for comparison. Bottom: The evolution of the radial temperature gradient in time similar to Fig.~\ref{fig:dyn-time-evol}. Here, we see that the innermost bump is a local maximum, producing a temperature gradient inversion, while the other two shadows produce slope changes in the profile.}
	\label{fig:inward-moving}
\end{figure}

We now present the results for the model with a surface density of $\Sigma_0 = 2000\,\text{g/cm}^2$ and a dust-to-gas ratio of $\edg = 0.01$, motivated in Sect.~\ref{sec:model-params}.
A higher surface density not only implies a higher dust content in the disk, with 10 times the amount of dust at a certain radius compared to the fiducial model but also a longer cooling timescale (in the hydrodynamic case the diffusion-regulated cooling timescale scales with $\Sigma^2\edg$, see also  Eq.~\eqref{eq:cooling-timescale}). Such a disk would be overall more optically thick than the \texttt{MFID} model, which would have an effect on the temperature structure.

Figure~\ref{fig:sigma-10} shows the temperature and the irradiation heating term for the hydrostatic and dynamical \texttt{MSIG2000} models. We see three temperature bumps at the end of the $30$ hydrostatic iterations, all within 10\,au of the disk. The time evolution of the staircase structure is clearer in Fig.~\ref{fig:inward-moving}, showing how the system readjusts from this static initial condition with three TWI bumps. In the hydrodynamic model, the inner bump continues to grow, altering the structure downstream: the staircases shift inward, and the first bump eventually grows sufficiently to merge with the second. The thermal response time for the disk is 100 times slower than the \texttt{MFID} model, and after about $7000$ inner orbits, we see a final state with two bumps (see also, bottom panel of Fig.~\ref{fig:sigma-10}, where the final state is plotted at $12000$ inner orbits). On more careful observation, we can also spot a smaller bump at $\sim20$\,au in the disk, with a relative temperature difference of $10\%$, less noticeable in the 2D profiles (Fig.~\ref{fig:sigma-10}) compared to the more prominent ones (see Sec.~\ref{sec:amplitude} for a discussion on the amplitude of the bumps in our models). The profiles of the radial temperature gradient in the bottom panel of Fig.~\ref{fig:inward-moving} show that the innermost bump in the hydrodynamic model causes a temperature gradient inversion i.e., a local maximum in the profile, something that was absent in the fiducial \texttt{MFID} model. The bumps downstream to it are more comparable, with the shadowing causing deviations in the slope of the temperature profile.

Overall, it is clear that the dust content in the disk has a significant effect on the staircase structure: a higher dust content correlates not only to more bumps that are more pronounced, but also show signs of inward motion. This  hints at more favorable conditions for the staircase structure in opaque disks, further enhanced by a driving inner bump.

Interestingly, our \texttt{MSIG2000} model shows inwardly moving waves similar to the predictions of the TWI, until it reaches a steady state after long-term evolution. This behavior could imply that optically thick disks provide more favorable conditions for the TWI itself, but identifying these conditions is beyond the scope of this work.

\section{Results III: Adding dust settling}
\label{sec:results-dust}

\begin{figure*}[ht]
        \centering
        \includegraphics[width=\linewidth]{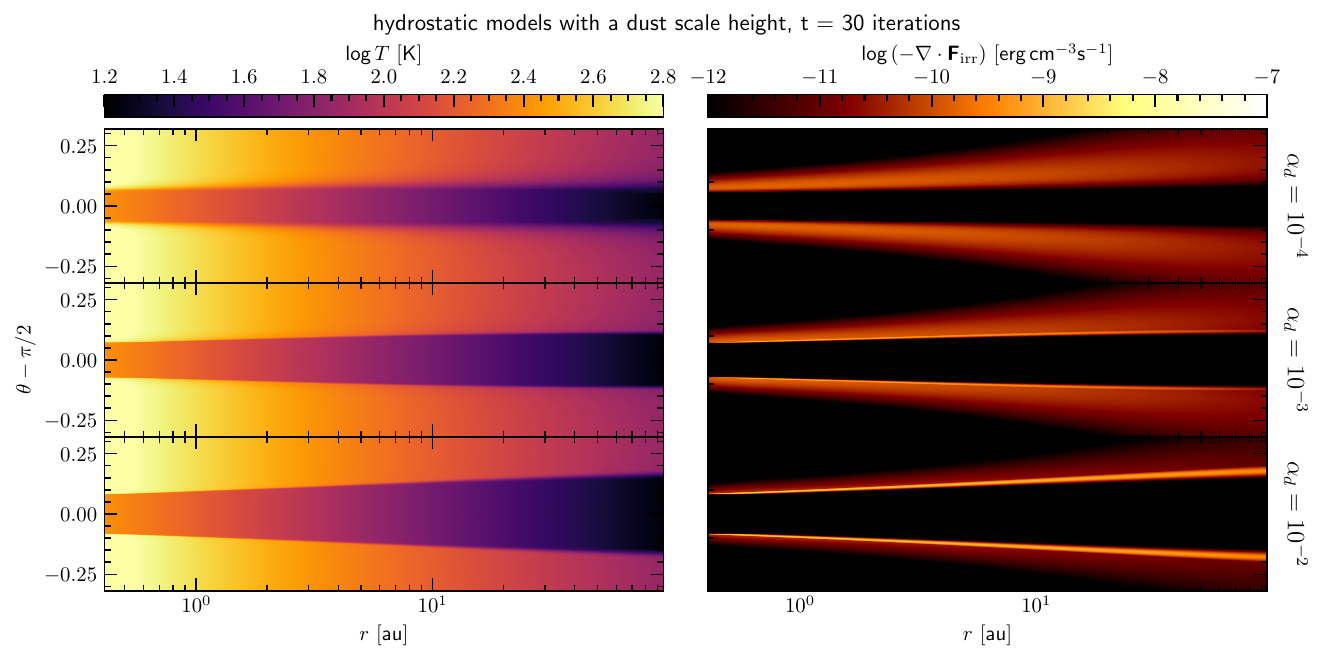}
        \caption{2D profiles of temperature and the irradiation heating term for the hydrostatic model with a dust tracer at $i = 30$ iterations, for diffusion coefficients $\alpha_\text{D} \in [0.0001, 0.001, 0.01]$. We see a very sharp irradiation surface but there are no staircases formed in any of these models.}
        \label{fig:dustscaleheight}
\end{figure*}
\label{sec:dust-scale}
With the dust content controlling radiation transport in the bulk of the disk via its opacity, modeling dust consistently is crucial to correctly capturing the shape of the thermal surface of the disk, as the latter determines the irradiation absorption surface. Dust settles at different heights and interacts with the gas through drag forces. In almost all previous models studying the thermal wave instability (including the results in our previous sections), there is an inherent assumption that the dust is perfectly coupled to the gas, and that the dust content scales with the gas mass through a constant dust-to-gas ratio. We present the first physically motivated hydrostatic model with dust by introducing a dust scale height within the hydrostatic radiation solver in \texttt{PLUTO} (see Sect.~\ref{sec:iterative-solver}). In the Epstein regime, the dust--gas drag force depends on the Stokes number $\St = t_\text{stop} \OmegaK$, which describes how well dust is coupled to the gas. Given a dust grain size $a$ and dust density $\rho_\text{d}$, the stopping time $t_\text{stop}$ is defined as
\begin{equation}
\label{eq:stokes}
t_\text{stop} = \frac{\rho_\text{d} a}{\rho \bm{v_\text{th}}},
\end{equation}
where $\bm{v_\text{th}} = \sqrt{8/\pi}\cs$ is the mean thermal velocity \citep{2024Birnstiel}. The vertical profile for the dust can be determined by the settling--diffusion equilibrium. A simple model usually imposes a Gaussian profile for the dust similar to the gas, although now with the dispersion being the "dust" scale height $H_\text{d}$ \citep[e.g., observations of IM Lup in][]{2008Pinte}. A more general approach involves modeling dust turbulence as a diffusive process \citep{1995Dubrulle,2004DullemondDominik,2012MuldersDominik}. A steady state solution for vertical settling and turbulent diffusion can be written in terms of the diffusion coefficient $D$ which quantifies diffusivity. For the simplest case, we utilize the solution assuming $D = \alpha_\text{d} \cs H$, where $\alpha_\text{d}$ is a dimensionless diffusion parameter. The dust density structure is then given by \citet{2009FromangNelson}
\begin{equation}
\label{eq:fromangnelson}
\rho_\text{d} =  \rho_\text{d, mid}\  \exp \left[- \frac{\text{St}_\text{mid}}{\alpha_\text{d}}\left(\exp\left(\frac{z^2}{2H^2}\right) - 1\right) - \frac{z^2}{2H^2}\right].
\end{equation}
We compute the dust density at the midplane from this formula at the end of every hydrostatic iteration once the new temperatures and gas densities are determined.
In our models, we consider a fixed grain size of $a = 10\,\mu$m and dust density $\rho_\text{d} = 3\,\text{g}\,\text{cm}^{-3}$. 

Figure~\ref{fig:dustscaleheight} shows 2D profiles of temperature and the irradiation source term for the dust models, for different diffusion parameters $\alpha_\text{D} \in [0.0001, 0.001, 0.01]$. We see a superheated dust layer and a very sharp irradiation surface for all cases, with dust settling more for a lower $\alpha_\text{D}$ \citep{2011Zsom}. For a given grain size, the dust--gas coupling is stronger where the gas density is higher. While the dust and gas are well coupled in the midplane, there is a sharp transition layer at the height where the dust is no longer well-coupled to the gas, and the dust density drops rapidly. Since the dust layer dictates the irradiation surface, this results in sharp layers where the majority of the irradiation from the star is absorbed \citep{1997ChiangGoldreich}.  
The height at which the radiation is absorbed is thus substantially closer to the midplane, an effect usually neglected in previous work on this topic. This affects the flaring angle of the disk, and hence the temperature structure. We see that staircases are not formed in any of these hydrostatic models with the dust scale height included.

While these models are already a first step to evolve dust in the disk, this iterative static settling--diffusion approach breaks down at a few gas scale heights. Our radiation transfer method accounts for small grain opacities through the precomputed opacity tables based on a grain size distribution, but the assumption of a constant $10\,\mu$m mass-carrying component, which would experience moderate settling is a simplification (see Sect.~\ref{sec:dust-scale} for a more detailed discussion on caveats of the existing dust models). To fully consistently model the transition of coupled-to-decoupled dust grains, one would have to model dust--gas interaction with drag and turbulent diffusion, along with the opacity feedback on radiation transport. We note that including this could affect the preliminary results with dust discussed in this section. Our future work will focus on this aspect in more detail, modeling dust as a pressureless fluid with the \texttt{PLUTO} code \citep[outlined in][]{2025Ziampras}.

\section{Discussion}
\label{sec:discussion}

In this section, we discuss some additional aspects of our models, such as the physical implications of the shadows and their observability, effect of the initial optical depth profile, the dust scale height, and the cooling timescales that could be important for the long-term evolution of shadows in the disk. We will also draw some qualitative comparisons to previous work on the thermal wave instability---to the linear theory and hydrostatic models of \citet{2021Wu}, \citet{2021Ueda}, and the hydrodynamic models of \citet{2022Pavlyuchenkov2D}, \citet{2022FuksmanKlahr}, and \citet{2024Kutra}. Finally, we highlight some caveats of our work and future directions.

\subsection{Physical implications of the bumps/shadows}
\label{sec:observable-shadow}
\begin{figure}[h]
	\centering
	\includegraphics[width=\columnwidth]{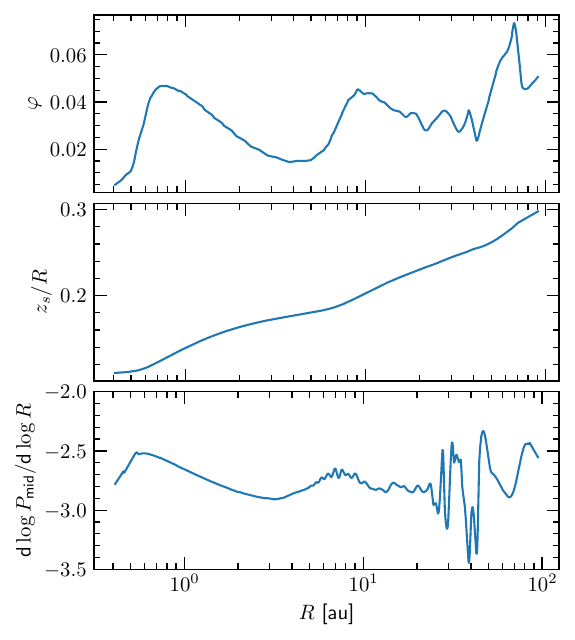}
	\caption{Top: Radial profile of the \referee{luminosity factor} $\varphi$ at the irradiation absorption surface showing decreased incidence of stellar rays in shadowed regions. Middle: Radial profiles of the surface aspect ratio $h$ at the $\tau_\star = 1$ surface. We see that the profiles deviate from a smooth power-law. Bottom: The midplane radial pressure gradient shows deviations in the pressure profile. We note the flaring angle $\varphi$ and the midplane pressure gradient $\text{d log}~P_{\text{mid}} / {\text{d log}~R}$ are smoothed with a rolling average over 20 cells to eliminate noise.}
	\label{fig:rad-quantities}
\end{figure}
    To understand if the staircase structure we see in our radiation hydrodynamic simulations can produce observable features, in addition to the relative midplane temperature profiles and the temperature gradients plotted in Figs.~\ref{fig:dyn-time-evol} and \ref{fig:inward-moving}, we provide further analysis here. The first two panels of Fig.~\ref{fig:rad-quantities} show the \referee{radial profiles of the luminosity factor $\varphi = \text{sin}\,\alpha$, where $\alpha$ is the flaring (grazing) angle and the height of the irradiation surface normalized to the local radius $z_s/R \text{ where } z_s = z~(\tau_\star=1)$. The flaring angle is the angle at which stellar rays hit the surface of the disk defined similar to the classical approach of \citet{1997ChiangGoldreich} and \citet{2000Dullemond}, where $\alpha = \arctan\left(\nicefrac{\text{d}z_s}{\text{d}R}\right) - \arctan\,(\nicefrac{z_s}{R})$ and the irradiation factor $\varphi$ is the fraction of luminosity intercepted by the disk surface, essentially approximating Eq.~\eqref{eq:irrad-flux} as
    \begin{equation}
        \label{eq:irrad-flux-approx}
        F_\star \approx \phi \frac{\Lstar}{4\pi r^2}.
    \end{equation}
    For small angles (i.e., razor-thin disks), $\varphi \simeq \alpha$}. We use the above mentioned quantities to trace whether the bumps/shadows produced in our models can affect the optical/NIR intensities producing observable signatures. We see that $\varphi$ decreases downstream of the bumps, blocking stellar rays to efficiently reach the shadows. This is also reflected by the non-uniform profiles of $z_s / R$, deviating from a smooth power-law profile. This provides evidence that the optical/NIR intensities are indeed reduced in the shadowed regions, but to also illustrate this further we provide synthetic scattered light images in the H Band (VLT/SPHERE) with RADMC-3D \citep{2012Dullemond} in Appendix.~\ref{sec:hband} showing that our models produce bright rings and subsequent dark shadows. 
We also plot the midplane pressure gradient for our models in the bottom panel of Fig.~\ref{fig:rad-quantities} to understand the implications of the shadows/bumps in our models in the mm continuum. We do not see the inversion of the pressure gradient \citep[i.e., dust traps, ][]{1972Whipple,2004Paardekooper} for any of our models, but we see deviations in the pressure gradient at the shadow locations. This implies that while the staircase structure due to the starlight-shadowing mechanism is not strong enough to trap dust, they could act as dust "traffic jams" that could slow down dust grain drift, and also act as seeds for secondary dust accumulation mechanisms \citep{2021JiangOrmel}. We highlight that the self-consistent modeling of dust--gas dynamics coupled with radiation transfer is required for more accurate constraints on observable signatures of shadows, which is why providing state-of-the-art synthetic observations both in NIR and the mm continuum is not the focus of this work.

\subsection{Vertical velocity profiles of the long-term evolved models}
\label{sec:long-term}
\begin{figure*}[ht]
        \centering
        \includegraphics[width=0.8\linewidth]{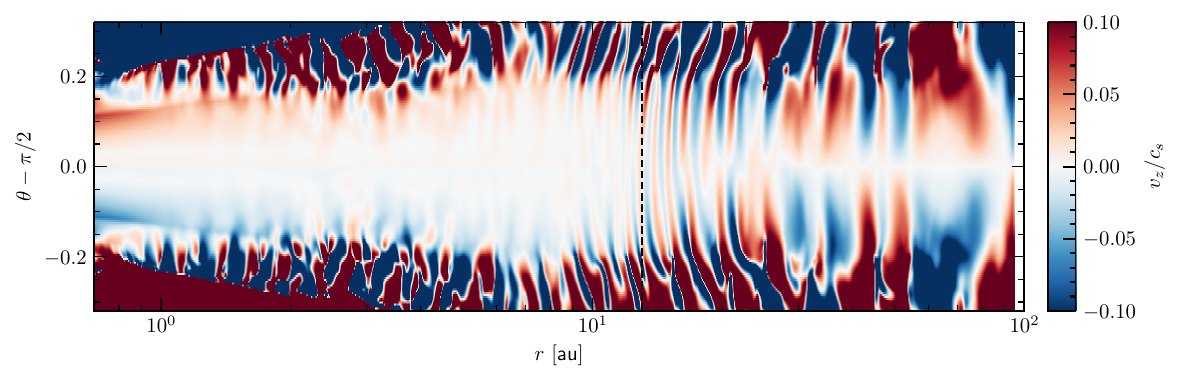}
        \caption{Vertical velocity profiles of the disk at $t = 10000$~orbits normalized to the sound speed. We see the development of the VSI modes in the outer disk. The dashed line indicates the location where VSI is expected according to the \citet{2015LinYoudin} criterion.}
        \label{fig:longterm}
\end{figure*}

As mentioned in Sect.~\ref{sec:long-term-1}, we ran the fiducial model \texttt{MFID} for several radiative diffusion timescales and saw long-lived staircases. Fig.~\ref{fig:longterm} shows the vertical velocity profile of the disk at $t = 10\,000$~orbits. In the outer disk ($\gtrsim 10\,$au), we see characteristic vertical motions extending from the top and bottom surfaces to the midplane (the so called finger modes) of the vertical shear instability \citep[VSI,][]{2013Nelson,2019Pfeil}. We also see some fully developed (corrugation) modes extending throughout the whole disk. The VSI is a hydrodynamic instability that could occur in outer protoplanetary disks where the cooling timescale is short, and can generate turbulence levels that match the ones observed in disks \citep{2022Dullemond,2025Villenave}. Although typical studies of the VSI require a higher resolution ($\sim16\,$cps) to capture saturated states, and even higher ($\sim50\,$cps) for accurate growth rates \citep{2020Flores}, one can still resolve the characteristic motions for grid sizes used in this study \citep{2023Ziampras}. A more detailed calculation of the cooling timescales in our model can be found in the Appendix~\ref{sec:cooling-timescales}. The VSI is expected to mix small grains in the disk efficiently \citep{2022Dullemond}. A combination of a VSI-active outer disk and a vertically settled dust layer in the inner regions could further result in a sharp radial transition of the height of the dust layer, affecting---or even promoting---structure due to starlight-driven shadowing.
This can only captured by models with self-consistent dust dynamics.

\subsection{Hydrodynamic modeling changes the picture: comparisons to linear theory and hydrostatic studies} 
\label{sec:linear-hyd}
Within the framework of linear analysis in the optically thick disk approximation, \citet{1999DAllesio} showed that the disk is stable to temperature perturbations for very long cooling times (e.g., in the inner disk) and \citet{2000Dullemond} found unstable thermal waves for short cooling times (e.g., in the outer disk). We find that temperature bumps/shadows are most likely between 1--30\,au, corresponding to intermediate-to-long cooling times \citep[similar to the findings of][]{2021Wu}. Time-dependent analysis by \citet{2008WatanabeLin} and \citet{2021Ueda} found shadows radially outwards up to $\sim$100\,au for dust-to-gas mass ratios up to $\edg = 0.001$. While both linear theory \citep{2000Dullemond}, and hydrostatic models \citep{2021Ueda,2021Wu,2022Okuzumi} found inwardly propagating waves, we find temperature bumps that grow and reach a quasi-steady configuration in our models, similar to models of \citet{2008WatanabeLin} with stabilized shadows for high mass accretion rates $\dot{M}$. While we see some inward motion of bumps in the \texttt{MSIG2000} model, they also reach a quasi-steady state after a few thousand orbits. 

We show that linear theory and vertically isothermal hydrostatic models do not accurately reflect the conditions in the disk. Consistent radial and vertical transport of radiation and advection indicate a more "quasi-steady" picture than a disk susceptible to a thermal wave or a self-shadowing ``instability''. While this non-linear ``quasi-steady'' state remains to be fully constrained, the natural formation of long-lived shadows in the disk is an interesting explanation for the origin of some of the ubiquitously observed substructures in protoplanetary disks.

\subsection{Comparisons to existing hydrodynamic models}
\label{sec:hydro}       
\citet{2024Kutra} performed hydrodynamic simulations with a simplified cooling model, where they relaxed the local temperature of the disk towards a "forcing temperature" calculated based on a vertical integration of the irradiation source term, over the thermal relaxation timescale. This restricts their model effectively to a vertically isothermal approximation, implying changes in the temperature at the surface are propagated downwards almost instantaneously likely favoring the formation of staircases \citep{2022FuksmanKlahr}.
They perform hydrodynamic simulations with gray irradiation, constant opacities for the dust, and find staircases forming in the disk that stall.

While our hydrodynamic models with frequency-dependent irradiation and the flux-limited diffusion method also predict a similar "quasi-steady" state, it is important to note key differences in physics. Solving 2D axisymmetric hydrodynamics with a simplified cooling model (one that is effectively vertically isothermal) constrains thermal evolution to being a purely radial process. Heating and cooling due to vertical compression, expansion, or radiative diffusion is affected by the imposed vertical temperature profile to which the system is forced to relax. We believe that this can cause unphysical propagation or even amplification of temperature perturbations, exaggerating the staircase structure \citep[to similar amplitudes like in 1.5D simulations, e.g.,][]{2021Ueda}. Moreover, the thermal relaxation timescale they use is vertically constant throughout the disk \citep[see Fig.~1 of][]{2024Kutra}, not capturing the transitions in optical depth from the disk surface to the midplane nor vertical heat diffusion. For context, their model matches closely to our \texttt{MSIG2000} model, but while their midplane reaches equilibrium after a few hundred orbits, the same takes several thousand orbits in our model.

\citet{2022Pavlyuchenkov2D} also performed hydrodynamic simulations with a diffusion method, one that handles radiation transport in the optically thick regime accurately, but is artifically too diffusive in the optically thin and transition regions. This results in excessively rapid cooling near the optically thin disk atmosphere, where dust densities---and therefore cooling rates, that are proportional to $\rho\edg\kappa_\mathrm{P}$---are extremely low. Our model uses a flux limiter to capture both regimes, and also includes frequency-dependent ray tracing which improves the treatment of the radiation field \citep{2013Kuiper}.  Their initial conditions differ from ours and more closely resemble a disk around a Herbig star with a luminosity of  $L = 5\,\mathrm{L}_\odot$. They include the inner rim, which when heated by a such bright star, puffs up and casts a wide shadow radially that extends outwards into the optically thin region, thereby suppressing other bumps in the domain. \citet{2024Kutra} showed that this shadow extended to about $20\,$au, and recovered the temperature bumps as the shadow was smaller when they used stellar properties similar to our work. They use a sparse grid resolution (especially in the vertical direction) which might prevent capturing the irradiation surface accurately, inhibiting the formation of shadows in the disk. 

\citet{2022FuksmanKlahr} used the M1 closure for radiation transport along with frequency-dependent irradiation. While in hydrostatic models we see that the M1 method produces temperature bumps that are more pronounced in amplitude than the ones we see with FLD, they always found that the thermal waves damp away when including hydrodynamics. This is in contrast to our findings, with the two shadows stabilizing for several hundreds of orbits in our simulations. We nevertheless compare our radial midplane temperature profiles and find that they match quite well (see Appendix~\ref{sec:fld-m1}). Our prescriptions of the inner disk and the initial optical depth differ, where we adopt a simpler prescription following Eq.~\eqref{eq:tau0} and \citet{2013Flock}. We discuss the implications of using different initial optical depth profiles in Sect.~\ref{sec:inner-disk} and the differences between our models in Appendix~\ref{sec:tau0-sampling}. 

The differences in the radiation methods among all of these existing hydrodynamic studies make it difficult to establish direct comparisons. This acts as motivation for future studies to compare the different methods in a systematic way to constrain the reliability of the irradiated disk studies. More complicated methods like the half-moment (HM) method \citep{2025Fuksman} and the discrete ordinate methods \citet{2021Jiang} could help resolve the discrepancy. 

\subsection{The amplitude of the bumps/shadows with FLD}
\label{sec:amplitude}
We see that the largest amplitude of the bumps (with the inner bump being the strongest) is about 40\% relative to the background power law. The shadows resolved with our FLD model are less steep compared to those resolved with other radiation methods. In comparison, bumps in the M1 hydrostatic models of \citet{2022FuksmanKlahr} showed amplitudes of order 50--60\% (see Fig.~\ref{fig:tempcomp}). The models of \citet{2024Kutra} show bumps that stand out by a factor of 2, more consistent with a 1D thermodynamic approach. With FLD, we also see that the bumps are not strong enough for a temperature gradient inversion (except for the inner bump in the \texttt{MSIG2000} model) as opposed to the above mentioned methods, where they are locations of local extrema.  These differences are expected with different radiation methods, details of are explored in code comparison works like \citet{2022Menon} and \citet{2022Thomas}.

\subsection{The inner disk structure and the initial optical depth}
\label{sec:inner-disk}
The structure and evolution of the inner rim of a protoplanetary disk is still an active field of research. The disk at the magnetospheric radius, studied by previous works like \citet{2001Dullemond} and in radiation hydrodynamical models like \citet{2019Flock, 2024Chrenko, 2025Flock}, is expected to feature a hot, puffed-up inner rim which casts a shadow upon the outer disk. Hence, there is still uncertainty about choosing the exact inner disk profile and whether the prescription for the optical depth of the inner disk $\tau_{\nu, 0}$ (see Eq.~\eqref{eq:tau0}) is realistic. Our prescription of the initial optical depth is similar to the one used in \citet{2013Flock} and could introduce a small hotspot in the inner disk (see Fig.~\ref{fig:tau0-comp}) which was not the case in the approach of \citet{2022FuksmanKlahr}. Nevertheless, this parameter cannot be constrained without more informed modeling of the inner disk temperature profile, and whether such a profile is smooth or structured. Studies further show that owing to material pileup at the inner edge of the dead zone, which can periodically trigger flares in MRI activity and accretion bursts, the inner disk can be hydrodynamically active and likely structured \citep{2006Vorobyov, 2024Cecil}. This can cause dynamical shadowing in the outer disk \citep[e.g., outwardly moving shadows along with static ones found in][]{2018Schneider}.

\subsection{Snowlines and opacity transitions}
\label{sec:snowlines}
The possibility that opacity transitions could trigger bumps, causing subsequent structure around it due to starlight-driven shadowing, could have physical and chemical implications on the structure in the outer disk close to snowlines, which are regions of high opacity shifts \citep{2007GaraudLin, 2011Oka,2015Bailli}. Some viscous models in the literature that include this show evidence of downstream shadowing \citep{2020Savvidou}. Moreover, we know that snowlines might not be thermally stable, and could be susceptible to the snowline instability \citep{2020Owen}. Dynamic changes of the snowline location, combined with secondary shadows downstream, might create ringed structures as in this work that could show variability in the disk. Secondary structures due to starlight-driven shadowing could also be important for planet formation models that invoke shadows around the snowline in the solar system \citep[for e.g., modeling the formation of Jupiter in the shadow cast by dust pileup near the water snowline in][]{2021Ohno}.

\subsection{The effect of dust}
\label{sec:gas-dust}
Our hydrostatic dust models in Sec.~\ref{sec:dust-scale} show very sharp irradiation absorption layers, and no temperature bumps regardless of the level of turbulent diffusion. This is in contrast with the models that take into account dust-settling using RADMC-3D in \citet{2021Wu}. The authors of this work show shadows due to the TWI extending out to $100$\,au, but their approach relied on an \emph{ad hoc} prescription of removing dust from the upper layers (above two scale heights). While this emulates a dust distribution with vertical settling, it does not explicitly account for turbulent diffusion, size dependence on settling and is not entirely physically self-consistent with the radiation transfer iterations. The qualitative impact of settling in determining the accurate irradiation absorption surface is robust from the dust models included in article. However, we note that relaxing the constant $10\,\mu$m grain size for the mass-carrying dust component could have an effect on the amplitude of the shadows, something worth exploring in the future.

Self-consistent hydrodynamic models of dust--gas interaction are an important next step towards understanding the shadows and the overall thermal structure of the disk. Our hydrostatic dust scale height models, even though they take into account a settling--diffusion equilibrium, can be improved further to account for the dust transition layer---that is, the height beyond where dust decouples dynamically from the gas. Moreover, the effect of dust backreaction and opacity feedback \citep[e.g.,][]{2024Krapp, 2025Ziampras} will be crucial to understanding the irradiation absorption surface, and subsequently any possible starlight-driven shadowing. For regions where dust--gas coupling is weak, decoupled dust temperatures with schemes like the three-temperature approach of \citet{2023Muley} might also become relevant.

\section{Summary}
\label{sec:summary}

We ran radiation hydrostatic and hydrodynamic simulations using frequency-dependent ray-traced irradiation and the flux-limited diffusion approach to study regions of the disk susceptible to starlight-driven shadowing and explore the existence of the thermal wave instability (TWI). We also present the first models that self-consistently capture the vertical structure of dust in the hydrostatic approximation. We outline our key findings below.
   \begin{enumerate}
      \item  We find that the best case scenario to recover starlight-driven shadows are in optically thick, slow cooling disks characterized by high dust-to-gas ratios in small grains $\edg=0.01$ or high surface densities (models \texttt{MFID}, \texttt{MSIG2000}). In the hydrostatic limit, we recover the TWI with thermal waves that grow and travel inwards.
      \item We see a strong effect of dynamics on the growth and final state of the shadows in the disk. We report that temperature bumps within $30\,$au of the disk grow, reach a quasi-steady state, and remain stable for several thousand orbits. We find the largest relative amplitude of 40\% for the bumps with respect to the background power law at the midplane. The inner bump is about $1.7-2\,$au wide, while the second bump is $10\,$au wide in the \texttt{MFID} model and $5\,$au wide in the \texttt{MSIG2000} model where we also see a third bump forming. 
      \item We do not see staircase structure propagating inwards at the timescales predicted by linear theory and 1D models, and hence do not recover the "instability" these models refer to, but a long-lived stable structure. 
      \item While the innermost bump in the disk could be influenced by the opacity transition at $250$\,K or the inner disk optical depth profile, we find that the consequent bumps could be natural consequences of an existing first bump, driven by the starlight-driven shadowing argument.
      \item The amplitude of the shadows is less pronounced than those recovered in the vertically isothermal limit \citep{2021Ueda,2024Kutra}. This could be a result of more accurate thermodynamic modeling, indicating the importance of accounting for vertical diffusion of temperature in any studies that rely on temperature transitions.
      \item We find that for lower dust-to-gas ratios $\edg\leq0.001$ the shadowing is suppressed. We only recover the innermost bump in the dynamic runs. 
      \item We present the first physically motivated hydrostatic models including dust settling. Our models with $10 \mu$m dust grains show a superheated dust layer without any staircases, indicating that dust impacts the irradiation absorption layer, and hence any subsequent shadowing. Further, the $\tau_\nu=1$ layer substantially moves closer to the midplane. We highlight that coupling the dust motion and dynamics to the opacity remains important. 
      \item Better constraints on the structure and optical depth of the inner disk---an important initial condition in our model, in light of recent self-consistent radiation hydrodynamic modeling of the inner rim casting a shadow on the outer, less turbulent disk \citep{2017Flock,2024Cecil,2025Flock} will help address the robustness of these starlight shadows in a more realistic scenario.
   \end{enumerate}
We note that a more detailed parametric study of different stellar properties, opacities, including snowlines etc., will help further constrain the limits of starlight-driven shadowing in protoplanetary disks. Furthermore, differences in results with various radiation methods need to be explored in detail, and would be decisive to our existing understanding of the temperature structure in protoplanetary disks. More refined methods with multi-group radiative transfer, or angle-dependent discrete ordinate methods, will constrain our models better.

It is also interesting to explore how observable these shadows are with synthetic observations. This mechanism could be a simple explanation to the rather common surface substructure like rings and gaps found in scattered light observations with VLT-SPHERE (and possibly in the future, with ELT facilities like MICADO and METIS). The subsequent impact on dust dynamics will also give us insight of how these shadows might manifest in the millimeter continuum, adding to our understanding of the observed substructures in protoplanetary disks.

\begin{acknowledgements}
The authors thank the anonymous referee for a thorough report, corrections and comments that improved this manuscript. PS thanks Satoshi Okuzumi and Thomas Pfeil for helpful comments. PS is grateful for the many discussions on steady states and equilibria with Siva Athreya. PS acknowledges the support of the German Science Foundation (DFG) through grant number 495235860 and is a Fellow of the International Max Planck Research School for Astronomy and Cosmic Physics at the University of Heidelberg (IMPRS-HD). MF acknowledges support from the European Research Council (ERC), under the European Union’s Horizon
2020 research and innovation program (grant agreement No. 757957). AZ and TB acknowledge funding from the European Union under the European Union’s Horizon Europe Research and Innovation Programme 101124282 (EARLYBIRD). Views and opinions expressed are those of the authors only. All plots in this paper were made with the Python library \texttt{matplotlib} \citep{2007Hunter}. Typesetting was expedited with the use of GitHub Copilot and ChatGPT, but without the use of any AI-generated text.
\end{acknowledgements}

\bibliographystyle{aa}
\bibliography{ii}

\begin{appendix}
\section{Comparing FLD to M1 moment transfer}
\label{sec:fld-m1}
\begin{figure}
	\centering
	\includegraphics[width=\columnwidth]{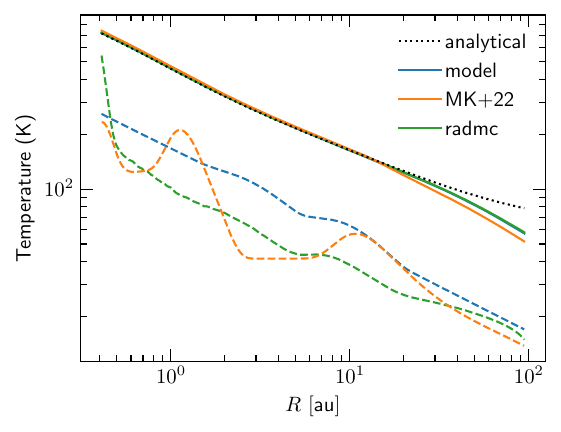}
	\caption{Comparison of 1D temperature profiles of the hydrostatic models at the end of the simulation for two different radiation methods --- flux-limited diffusion (FLD, used in this work) and moment transfer \citep[M1, ][]{2022FuksmanKlahr}. Overall we see very good agreement with temperatures with FLD and M1, with the optically thin atmosphere being captured quite well. The black dotted line shows the analytical solution for the temperature profile in the optically thin atmosphere \citep{1997ChiangGoldreich}. The midplane temperatures are also in good agreement, with some minor deviations. The M1 method shows higher amplitude bumps than the FLD method. We also include the RADMC-3D calculations corresponding to our FLD model, these correspond to the green lines.}
	\label{fig:tempcomp}
\end{figure}
We compare the final structures from the hydrostatic models of \citet{2022FuksmanKlahr} who use the M1 moment transfer method, as opposed to the flux-limited diffusion approach we use in this work, in Fig.~\ref{fig:tempcomp}. The parameters used for the model corresponds to the \texttt{ML1} run (see Table~\ref{tab:modelparams}). The temperature profiles at the optically thin atmosphere match between the models very well, and to the analytical solution by \citet{1997ChiangGoldreich} and the corresponding RADMC-3D calculations (green line in Fig.~\ref{fig:tempcomp}). The background power-law of the midplane temperatures are also in good agreement, with some minor deviations given the different radiation treatments. Both radiation methods show the formation of thermal waves. However, with the M1 method, we see that the observed bumps are of higher amplitude than the ones we see with FLD. 

Both the M1 and the FLD methods overestimate the temperature at the midplane, which is expected to be lower using Monte Carlo methods \citep{2002Dullemond,2025Krieger,2025Fuksman,2025Pavlyuchenkov}. This is also clear in in Fig.~\ref{fig:tempcomp} by the comparison of the FLD model (blue dotted line) to the corresponding RADMC-3D calculation (green dotted line). \citet{2024Kutra} argued that this casts doubt on the validity of the results in \citet{2022FuksmanKlahr}, suggesting that this effect may preclude the formation of a temperature staircase. However, despite this overestimation, both FLD and M1 models reproduce shadowed structures in hydrostatic simulations, and we find a parameter space where shadows persist in hydrodynamic models with FLD.
Thus, we find the argument presented in \citet{2024Kutra} to lack strong physical motivation. Instead, the presence or absence of a staircase is sensitive to the assumed dust content, the production of initial bumps, and the correct treatment of diffusive cooling timescales resulting from the balance of stellar irradiation and thermal cooling.

\section{H Band scattered light synthetic observation}
\label{sec:hband}
\begin{figure}
	\centering
	\includegraphics[width=\columnwidth]{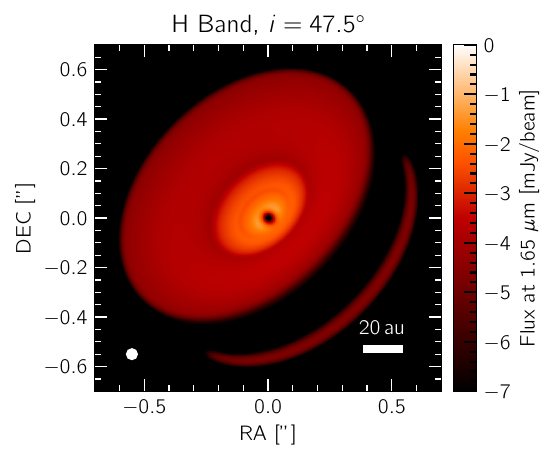}
	\caption{Synthetic scattered light images in the H Band of the final disk state in the fiducial model \texttt{MFID}. We chose a disk at a distance of $158\,$pc and an inclination of $47.5\deg$ similar to IM Lup \citep{2018Avenhaus}. The square in the bottom left denotes our beam size with a FWHM of $40''\times40''$.}
	\label{fig:hband}
\end{figure}
We post-processed the final temperature state from our fiducial radiation hydrodynamic model \texttt{MFID} using the radiation transfer code RADMC-3D \citep{2012Dullemond} to produce synthetic scattered light images that could be observed using VLT/SPHERE in the H Band (1.65\,$\mu$m). We do not recompute the temperatures with \texttt{radmc}~\texttt{mctherm} but we recalculated the dust opacities with \texttt{optool} \citep{2021Dominik} with the same parameters as the ones used in this article \citep{2020AKriegerWolf} to also include the scattering components. We then convolve the image using a Gaussian filter with a beam size of $40''\times40''$, and add a coronograph of the same size around the star. Fig.~\ref{fig:hband} shows the scattered light image with the two starlight-driven bumps/shadows in our model having an imprint in the observables.

\section{Thermal relaxation timescale}
\label{sec:cooling-timescales}
\begin{figure}
        \centering\includegraphics[width=\columnwidth]{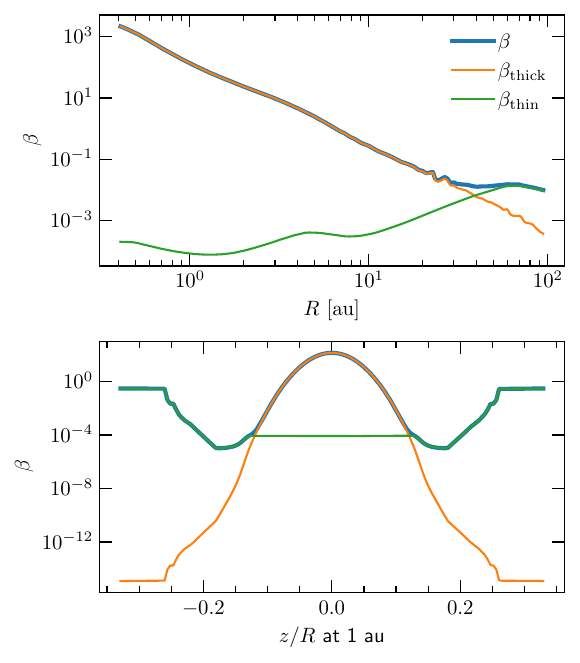}
        \caption{Top: Radial profile of the cooling timescale $\beta$ for the disk.  The regions of the disk where we find starlight-driven shadows lie in the intermediate cooling timescale regime ($1-30\,$au). Bottom: Vertical profile of the cooling timescale $\beta$ at $R = 1$~au.  We see that for most of the disk (except for the atmosphere, and outermost regions), optically thick cooling is more dominant.}
        \label{fig:cooling-time}
\end{figure}

Previous work has supported the idea that the ratio between the cooling timescale $t_\text{cool}$ and the dynamical timescale ($\approx\OmegaK^{-1}$) is important in determining the formation and propagation of thermal waves \citep[e.g.,][]{2000Dullemond,2022FuksmanKlahr}. We therefore compute the cooling timescale to both assess its effect on the dynamics of the TWI as well as to estimate the radiative diffusion timescale between hydrostatic iterations in our models. For well-coupled dust and gas, the cooling parameter $\beta = t_\text{cool} \OmegaK$ can be written as a sum of optically thick and thin cooling components in the flux-limited diffusion approach \citep[following][]{2017Flock, 2024bZiampras},
\begin{equation}
\label{eq:cooling-timescale}
\beta = \frac{\OmegaK}{\eta} \left(H^2 + \frac{l^2}{3}\right), \quad \eta = \frac{16 \sigsb T^3}{3 \kappa_P \rho^2 \cv}, \quad l = \frac{1}{\kappa_P \rho} 
\end{equation}
where $\sigsb$ is the Stefan-Boltzmann constant. We plot $\beta$ radially and at a vertical slice corresponding to $R = 1$\,au in Fig.~\ref{fig:cooling-time}. Most of our radial computational domain is dominated by the optically thick component ($\propto H^2$) until up to $R \sim 50$\,au. We find starlight-driven staircases between 1--30\,au, in regions corresponding to long-to-intermediate cooling timescales radially. Cooling is inefficient in the inner parts where the cooling parameter is as high as $\beta = 10^3$ and then it decreases reaching up to $\beta = 0.1$ at $\sim15$\,au. This indicates that the outer part of the disk is susceptible to the vertical shear instability \citep{2013Nelson, 2019Pfeil, 2020Flock}, as also supported by our study (see Fig.\ref{fig:longterm}).
In the inner disk and along the vertical direction, the region around the midplane remains optically thick up to about four scale heights ($|z|/r\lesssim0.12$), with a rapidly-cooling optically thin atmosphere where $\beta\lesssim10^{-4}$. The spikes in the vertical profiles of the cooling timescale correspond to the numerical artifact in the temperature (see Fig.~\ref{fig:dyn-time-evol}) and the flat profile in the the upper disk boundaries corresponds to the density floor.

\section{The effect of $\tau_{\nu, 0}$ on the irradiation source term}
\label{sec:tau0-sampling}
In Sect.~\ref{sec:inner-disk} we discussed the effect of the optical depth profile $\tau_{\nu, 0}$ at the inner radial end of our domain, given that it is an important initial condition that sets our disk model. We do not fully know the exact structure of the inner disk in reality. Starting out with shadows in the inner disk suggests that the disk outside of it could be subject to non-uniform heating. This could further help starlight-driven shadowing. In this section, we compare the effect of two different profiles of $\tau_{\nu, 0}$ on the irradiation source term.

\begin{figure}
	\centering
	\includegraphics[width=\columnwidth]{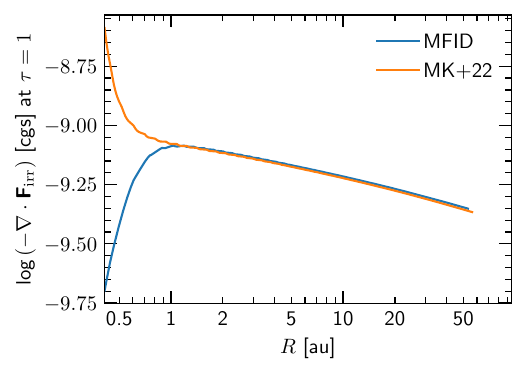}
	\caption{Comparison of the irradiation source term sampled at the $\tau = 1$ surface for two different profiles for the initial optical depth $\tau_{\nu, 0}$. The blue line shows the profile used in this work while the orange line shows the profile used in \citet{2022FuksmanKlahr}. We see that they match very well outside $1\,$au but differ inwards. While  \citet{2022FuksmanKlahr} use a radially decreasing, smooth profile, we see that our prescription can create little non-uniformities in heating that could influence the innermost bump.}
	\label{fig:tau0-comp}
\end{figure}

We compare the profile used in this work (Eq.~\eqref{eq:tau0}) with the one used in \citet{2022FuksmanKlahr}. Fig.~\ref{fig:tau0-comp} shows the irradiation source term sampled at the $\tau = 1$ surface for both cases. We find very good agreement between the two prescriptions through most of the disk outwards of 1\,au. In the innermost parts, we note that the orange line \citep[used by][]{2022FuksmanKlahr} is a smooth, radially decreasing quantity, while there is a little hotspot in the profile we use. This could influence the growth of the innermost bump in our model. Better models that motivate the inner disk structure from self-consistent radiation hydrodynamic models will help constrain this initial condition better. We nevertheless stress that the presence of a second bump (e.g., at 10\,au in Fig.~\ref{fig:dyn-time-evol}) is unaffected by this.
\end{appendix}

\end{document}